\documentclass[a4paper]{llncs}

\usepackage{amssymb}
\usepackage{graphicx}

\usepackage{url}
\urldef{\mailsa}\path|avni.verma@research.iiit.ac.in,|
\urldef{\mailsb}\path|kamal@iiit.ac.in|
\urldef{\mailsc}\path||

\usepackage{enumerate}
\usepackage{fixltx2e}
\usepackage{hyperref}
\usepackage{multirow}
\usepackage{float}
\usepackage{tikz}
\usepackage{tikz-qtree}
\usepackage{array}
\usepackage{longtable}
\usepackage{caption}
\usepackage{amsmath}
\usepackage{amssymb}
\usepackage{algorithm,algpseudocode}
\usepackage{placeins}
\usepackage{float}
\usepackage{subfig}
\newtheorem{thm}{Theorem}
\newtheorem{lem}[thm]{Lemma}
\newtheorem{conj}{Conjecture}

\newcolumntype{C}[1]{>{\centering\let\newline\\\arraybackslash\hspace{0pt}}m{#1}}

\newcommand\floor[1]{\lfloor#1\rfloor}

\begin{document}

\mainmatter  

\title{Design and Evaluation of Alternate Enumeration Techniques for Subset Sum Problem}


%
%
%


%
%


\author{Avni Verma and Kamalakar Karlapalem \\ \mailsa \mailsb }
\institute{
{Data Science and Analytics Center (DSAC)\\ International Institute of Information Technology, Hyderabad, India}}

\maketitle

\begin{abstract}
The subset sum problem, also referred as SSP, is a NP-Hard computational problem. SSP has its applications in broad domains like cryptography, number theory, operation research and complexity theory. The most famous algorithm for solving SSP is Backtracking Algorithm which has exponential time complexity. Therefore, our goal is to design and develop better alternate enumeration techniques for faster generation of SSP solutions. Given the set of first $n$ natural numbers which is denoted by $X_{n}$ and a target sum $S$, we propose various alternate enumeration techniques which find all the subsets of $X_{n}$ that add up to sum $S$.

In this paper, we present the mathematics behind this exponential problem. We analyze the distribution of power set of $X_{n}$ and present formulas which show definite patterns and relations among these subsets. We introduce three major distributions for power set of $X_{n}$: Sum Distribution, Length-Sum Distribution and Element Distribution. These distributions are prepossessing procedures for various alternate enumeration techniques for solving SSP. We propose novel algorithms: Subset Generation using Sum Distribution, Subset Generation using Length-Sum Distribution, Basic Bucket Algorithm, Maximum and Minimum Frequency Driven Bucket Algorithms and Local Search using Maximal and Minimal Subsets for enumerating SSP.

We compare the performance of these approaches against the traditional backtracking algorithm. The efficiency and effectiveness of these algorithms are presented with the help of these experimental results. Furthermore, we studied the over solution set of subsets generated by various algorithms to get the complete solution for subset sum problem.  Finally, we present a conjecture about upper bound on the number of subsets that has to be enumerated to get all solutions for Subset Sum Problem.
\end{abstract}

\section{Introduction}
\label{Ch1:Introduction}
In SSP, we consider a set of $n$ positive integers  stored in set $X$ and a target sum $S$. $X = \{x_{1}, x_{2} \ldots x_{n}\}$. Traditionally, there are two definitions for SSP which are described below:
\begin{enumerate}
\item Version $1$: Given a set $X$ containing positive integers and a target sum $S$, is there a subset which sum upto $S$? This is a NP-Complete problem.

For example, given $X = \{5, 4, 9, 11\}$ and $S = 9$, the solution to this problem is \textit{true}. There are many ways to solve this problem and it depends on the size and values of $X$ and $S$. The brute force algorithm iterates through all possibilities and takes $\mathcal{O}(2^{n} \times n)$ time for execution. For smaller size and values of $X$ and $S$, SSP can be solved in polynomial time by using dynamic programming with time complexity $\mathcal{O}(n \times S)$ \cite{Authors03Wiki}.

\item Version $2$: Given a set $X$ containing positive integers and a target sum $S$, find a subset which can sum up to $S$. This is a NP-Hard problem.
 
For $X = \{5, 4, 9, 11\}$ and $S = 9$, the solution to above problem is either $\{5, 4\}$ or $\{9\}$. This is a exponential time taking  problem which can be solved in $\mathcal{O}(2^{n} \times n)$ time by using brute force. This method requires $\mathcal{O}(n)$ storage space to store the required result. This version of SSP does not have any known polynomial time algorithm.
\end{enumerate}

In this paper, we extend the traditional SSP (Version $2$) and design various alternate enumeration techniques. Instead of finding one subset with target sum, we find all possible solutions of SSP. Therefore, for $X = \{5, 4, 9, 11\}$ and $S = 9$, solutions to our version of SSP are $\{5, 4\}$ and $\{9\}$. We further confine and refine our problem domain by considering first $n$ natural numbers as set $X$. There are many advantages for selecting this problem domain. It simplifies the problem statement, avoids duplication and since sum of first $n$ natural number is $\frac{n(n+1)}{2}$, by selecting $X = \{1, 2 \ldots n\}$ we restrict target sum between $1$ and $\frac{n(n+1)}{2}$, $S \in [1, \frac{n(n+1)}{2}]$. The efforts to solve Subset Sum Problem are required to get subset queries in relational databases \cite{valluri2004subset}. Before describing the formulation of our problem in detail we explore the research work conducted in field of SSP.

\section{Related Work}
\label{Related Work}
The Subset Sum Problem has been studied very widely. It has a standard $\mathcal{O}(nu)$ pseudo-polynomial time dynamic programming algorithm \cite{Authors01} which is taught in elementary algorithms class. Additionally, there are a number of other algorithms in the literature, including an FPTAS \cite{Authors02}, an exact algorithm with space and time trade offs \cite{Authors03}, a polynomial time algorithm for most low density sums \cite{Authors04}, and a number of more specialized pseudo-polynomial time algorithms with various properties \cite{Authors05} \cite{Authors06} \cite{Authors07} \cite{Authors08}. 

There is also another variant of Subset Sum Problem which allows the elements in $X$ to be used any number of times in the sums. Overall, dynamic programming is expected to be most efficient for very dense instances, while backtracking is expected to be most efficient for sparse instances of Subset Sum Problem. In different versions of the SSP, the input set may or may not contain duplicate values, and the problem can also be expressed as an optimization problem. 

In \cite{Authors09}, the authors have introduced a new faster pseudo-polynomial time algorithm for the Subset Sum problem to decide if there exists a subset of a given set $S$ whose elements sum to a target number $t$. Their proposed algorithm runs in  $\mathcal{O}(\sqrt{n}t)$  time, where $n$ is the size of set $S$. Their approach is based on a fast Minkowski sum calculation that exploits the structure of subset sums of small intervals. 

Despite the apparent simplicity of the problem statement, to date there has been modest progress on exact algorithms\cite{Authors11} for Subset Sum Problem. Indeed, from a worst-case performance perspective the fastest known algorithm runs in $\mathcal{O}(2^{\frac{n}{2}})$ time and dates to the 1974 work of Horowitz and Sahni\cite{Authors12}. Improving the worst-case running time is a well-established open problem\cite{Authors13}. 

In \cite{Authors18}, the authors present a randomized algorithm. They consider positive integers and a target sum but instead of fidning all subsets of target sum, the solution is bounded by $B$ concentration. The main result of this algorithm is that all instances without strong additive structure (without exponential concentration of sums) can be solved faster than the Horowitz-Sahni time bound $\mathcal{O}(2^{\frac{n}{2}}$) \cite{Authors12}. They have also shown a quantitative claim to show or prove it. Complexity of this randomized algorithms is $\mathcal{O}(2^{0.3399n}B^{4})$. 

Beier and Vocking \cite{Authors14} presented an expected polynomial time algorithm for solving random knapsack instances. Knapsack and subset sum have similarities, but the random instances considered there are quite different from ours, and this leads to the development of quite a different approach.
Subset sum problem is also closely related to the classical number theory study of determining partitions. In \cite{Authors15} Hardy and Wright provide generating functions but is limited due to lack of computational scheme for generating such partitions. A survey of algorithms for the different variations of the knapsack problem is given in \cite{Authors14}. Much of the early work in the knapsack problem was done by Gilmore and Gomory \cite{Authors16} \cite{Authors17}.

However, there is very little work done on enumeration techniques for subset sum problem, which we addressed in this work. We have developed different algorithms for alternate enumerations techniques for subset sum problem and have compared their performance.

\section{Formulation for Subset Sum Problem}
\label{Ch2sec:Formulation}
\noindent The following set of information is used for presenting the exponential aspect and solution of alternate enumeration techniques of SSP:
\begin{enumerate}
\item A set of first $n$ natural numbers. $X_{n} = \{1, 2, 3 \ldots n\}$ where $n$ is a positive integer. The set $X_{n}$ is also known as the \textit{Universal set}. This is our problem domain. The cardinality of the set $X_{n}$ is $n$. $|X_{n}| = n$

\item A set of all subsets of $X_{n}$ is $\mathcal P \left({X_{n}}\right) = \{\phi, \{1\},\{2\} \ldots \{1,2 \ldots n\}\}$. It is also known as power set. The empty set is denoted as $\phi$ or $\{\}$ or the null set. In this paper, we use $\phi$ for the representation. $|\mathcal P \left({X_{n}}\right)| = a = 2^{n}$

\item $maxSum(n)$ is the sum of all elements of the universal set $X_{n}$. This is the maximum possible sum for any element of $\mathcal P \left({X_{n}}\right)$.\\
\hphantom{abc} $maxSum(n) = b = (1+2+3 \ldots n) = \frac{n(n+1)}{2}$.\\
\hphantom{abc} $Sum(A) \leq maxSum(n) = \frac{n(n+1)}{2} \  \forall A \in \mathcal P \left({X_{n}}\right)$

\item $Sum(A)$ is the sum of all elements of a set $A$ where $A$ belongs to power sets of $X_{n}$, $A \in \mathcal P \left({X_{n}}\right)$.
\begin{itemize}
\item We assume sum of all elements of $\phi$ as 0, $Sum(\phi) = 0$. \item The range of $Sum(A)$ is $[0, \frac{n(n+1)}{2}]$.
\item The minimum possible sum for $A$, where $A \in \mathcal P \left({X_{n}}\right)$, is denoted as $minSum(n)$.
\end{itemize}

\item $midSum(n)$ is the mid point of the range of $Sum(A)$ where $A \in \mathcal P \left({X_{n}}\right)$. Since, the maximum possible sum for power sets of $X_{n}$, $\mathcal P \left({X_{n}}\right)$ is $\frac{n(n+1)}{2}$ and minimum possible sum is $0$, $midSum(n) = \frac{minSum(n) + maxSum(n)}{2} = \frac{0 + \frac{n(n+1)}{2}}{2}$  \\
\hphantom{abc} $midSum = d = \frac{(1+2+3 \ldots n)}{2} = \frac{n(n+1)}{4}$\\
\hphantom{abc} For simpler calculations, we consider $midSum$ as the largest integer less than or equal to the mid point, $floor(midSum(n)) =  \floor{\frac{n(n+1)}{4}}$.

\item $maxLength(n)$ is the count of all elements of the universal set $X_{n}$. This is the maximum possible length for any element of $\mathcal P \left({X_{n}}\right)$.
\begin{itemize}
\item Therefore, $maxLength(n)$ is equal to the cardinality of set $X_{n}$, defined in point-$1$.
\item $maxLength(n) = |X_{n}| = |\{1, 2 \ldots n\}| = n$  
\end{itemize}

\item $Len(A)$ is the count of all elements of a set $A$ where $A$ belongs to power sets of $X_{n}$, $A \in \mathcal P \left({X_{n}}\right)$.
\begin{itemize}
\item The range of $Len(A)$ is from $1$ to $n$, $Len(A) \in [1, n]$. 
\item We consider, count of all elements of subset $\phi$ as 1. $Len(\phi) = 1$. 
\item Therefore, the range of $Len(A)$ is from $1$ to $n$. $Len(A) \in [1, n]$.
\item The minimum possible length for $A$, where $A \in \mathcal P \left({X_{n}}\right)$, is denoted as $minLen(n)$.
\end{itemize}

\item $minSum(n, l)$ is the sum of a subset $A$ where $A \in \mathcal P \left({X_{n}}\right)$ with $Len(A) = l$. $A$ is the subset of length $l$ with minimum possible sum. Subset of length $l$ with minimum possible sum contains first $l$ smallest natural numbers. Therefore, minimum possible subset of length $l$ is $A = \{1, 2 \ldots l\}$.\\
\hphantom{abc} $minSum(n, l) = (1+2+\ldots+l) = \frac{l(l+1)}{2}$

\item $maxSum(n, l)$ is the sum of a subset $A$ where $A \in \mathcal P \left({X_{n}}\right)$ and $Len(A) = l$. $A$ is the subset of length $l$ with maximum possible sum. Subset of length $l$ with maximum possible sum will contain $l$ largest natural numbers decreasing from $n$.
\begin{itemize}
\item Maximum possible subset of length $l$ is $A$, $A = \{n, n-1 \ldots n-(l-1)\}$.
\item $maxSum(n, l) = (n+(n-1)+\ldots+n-l+1) = n \times l-\frac{l-1(l-1+1)}{2}$
\item $maxSum(n, l) = \frac{l(2n-l+1)}{2}$
\end{itemize}
\end{enumerate} 

\section{Distribution Formulae}
\label{Distribution Formulae}
We have analyzed the distribution of $\mathcal P \left({X_{n}}\right)$ over sum, length and count of individual elements. We present distribution formulas and algorithms, along with example, which show definite patterns and relations among these subsets.

In table \ref{Ch2Table:FormulaSummary}, we briefly present the formula, definition, meaning, values and assumptions of all distributions which are required for design and evaluation of alternate enumeration techniques for SSP. Cardinality of a set is the number of elements of the set. These distributions are prepossessing procedures which are required for presenting our novel alternate enumeration techniques for solving SSP. The formulae are the notation developed in Section-\ref{Ch2sec:Formulation}. In Table \ref{Ch2Table:FormulaSummary}, $b$ denotes the maximum possible sum for any element of $\mathcal P \left({X_{n}}\right)$, $b = \frac{n(n+1)}{2}$.

\captionsetup{width=\linewidth}
\begin{longtable}{| p{.13\textwidth} | p{.29\textwidth}| p{.29\textwidth} | p{.29\textwidth}|} 
\hline
Distribution & Formula  & Meaning & Value/Assumption\\
\hline\hline
$SD$ \newline Sum-Distribution & A 2D matrix with cardinality $n \times b$, where $|X_{n}| = n$ and $b = \frac{n(n+1)}{2}$. & $SD[n][S]$ represents the count of all the subsets belonging to $\mathcal P \left({X_{n}}\right)$ with sum $S$. \newline Every row, $SD[n]$, is the sum distribution for all subsets of $X_{n}$ where sum is $S$. & In this thesis, the empty set $\phi$ is counted once while calculating the sum distribution, $SD[n][0] = 1$. \\ \hline
$LD$ \newline Length-Sum-Distribution &  A 3D matrix of cardinality $n \times b \times n$, where $|X_{n}| = n$ and $b = \frac{n(n+1)}{2}$. & $LD[n][S][l]$ represents the count of all the subsets belonging to $\mathcal P \left({X_{n}}\right)$ with sum $S$ and length $l$. \newline Every column of this matrix, $LD[n][S][l']$, where $\forall l' \in [0, n]$, is the length distribution for all subsets of $X_{n}$ with sum $S$. & Extending the previous assumptions we get, \newline $LD[n][S][0]=1, \  \forall S \in [0, b]$ \newline $LD[n][0][l]=1, \  \forall l \in [1, n]$\\ \hline
$ED$ \newline Element-Distribution & A 3D matrix of cardinality $n \times b \times n$, where $|X_{n}| = n$ and $b = \frac{n(n+1)}{2}$. & $ED[n][S][e]$ represents the count element $e$ in all the subsets belonging to $\mathcal P \left({X_{n}}\right)$ with sum $S$. \newline Every row, $ED[n][S]$, is the element distribution for all subsets of $X_{n}$ with sum $S$. & In this thesis, we assume the count of element-$\phi$ in all subsets of $\mathcal P \left({X_{n}}\right)$ as $0$. \newline $ED[n][S][0] = 0, \ \forall S \in [0, b]$ \newline \newline A zero-sum is achieved only by subset $\phi$. \newline $ED[n][0][e] = 0, \ \forall e \in [0, n]$.\\ \hline
\caption{Formula, definition, meaning, values and assumptions of all distributions which are required for design and evaluation of alternate enumeration techniques for SSP. First column denotes the distribution name, second and third column define the formula, definition and concept behind every distribution and fourth column states all the assumptions.}
\label{Ch2Table:FormulaSummary}
\end{longtable}

\subsection{Sum Distribution}
\label{Ch2:SumDistribution}
In sum distribution, also referred as $SD$, we find the number of subsets which sum up to a certain integer $S$, where $X_{n} = \{1,2,3 \ldots n\}$ and $S \in [0, \frac{n(n+1)}{2}]$. It is represented as $SD[n][S]$. Equation \ref{Ch2Eq1} establishes the formula for the sum distribution. Before counting the subsets of a particular sum, we initialize the count as zero, $\forall n, S \ SD[n][S] = 0$. Following are the base cases for sum distribution $(SD[n][S])$:
\begin{enumerate}
\item For $n = 0$ and $S = 0$, the corresponding subset is $\phi$. Since, zero-sum ($Sum=0$) can be achieved only with subset $\phi$ and $Sum(\phi)$ is assumed to be $0$, as defined in Section \ref{Ch2sec:Formulation}, the count of occurrence of $\phi$-subset in $P(X_{0})$ is taken as 1. Therefore, $SD[0][0] = 1$.
\item $\forall i \in [1,n]$ and $S = 0$, $SD[i][0] = 1$. Since, zero-sum ($Sum=0$) can be achieved only with subset $\phi$, the count of occurrence of $\phi$-subset in $P(X_{n})$ is taken as 1. Therefore, $SD[n][0] = 1$.
\item $SD[i][j] = 0$, if $i < 0$ or $j < 0$.
\end{enumerate}

\begin{equation}
\label{Ch2Eq1}
    SD[n][S] = \begin{cases}
               1               & (S = 0) \ \text{or} \ (n=1) \\
               0 			   & (n = 0) \\
               SD[n-1][S]      & 0 < S < n\\
               SD[n-1][S] + SD[n-1][S-n] & n \leq S \leq \floor{\frac{n(n+1)}{4}}  \\
               SD[n][maxSum(n) - S] & \floor{\frac{n(n+1)}{4}} < S \leq maxSum(n) = \frac{n(n+1)}{2} \\
               0 & \text{otherwise}
           \end{cases}
\end{equation}
\noindent Similar to Element Distribution (Section \ref{Ch3ElementDistribution}), we can give uniqueness and correctness proof of Sum Distribution.

\subsection{Length-Sum Distribution}
\label{Ch2:LengthSumDistribution1}
In length-sum distribution, we find the number of subsets of $X_{n}$ of length $l$ which sum up to $S$ where $S \in [0, maxSum(n)]$, $maxSum(n) = \frac{n(n+1)}{2}$ and $l \in [0,n]$. Table \ref{Ch2TableLengthSumDisBaseCases} presents the bases cases for Length-Sum Distribution.

\begin{equation}
\label{Ch2Eq12}
    LD[n][S][l] = \begin{cases}
    		   1 & l = 0  \  \text{and} \ S = 0 \\
               LD[n-1][S][l]               & 1 \leq l \leq \floor{\frac{n}{2}} \  \text{and} \  0 \leq S < n \\
               LD[n-1][S][l] + LD[n-1][S-n][l-1]   & 1 \leq l \leq \floor{\frac{n}{2}} \  \text{and} \ n \leq S \leq \frac{n(n+1)}{2}\\
               LD[n][maxSum(n)-S][n-l]    & \floor{\frac{n}{2}} < l \leq n\\
               0 & \text{otherwise} \\
           \end{cases}
\end{equation}
\noindent Similar to Element Distribution (Section \ref{Ch3ElementDistribution}), we can give uniqueness and correctness proof of Length-Sum Distribution.

\begin{table}
\footnotesize
\begin{tabular}[t]{|c|c|c|c|}
\hline
\parbox[t]{0.5in}{\centering Values of $l$ \par for $n=0$ } & Subset & Sum of the Subset & \parbox[t]{1in}{\centering No. of Subsets / \par Length Distribution}\\
\hline\hline
l=0 & \{$\phi$\} & 0 & 1 \\
\hline
\end{tabular}
\hfill
\hspace{3em}
\begin{tabular}[t]{|c|c|c|c|}
\hline
\parbox[t]{0.5in}{\centering Values of $l$ \par for $n=1$ } & Subset & Sum of the Subset & \parbox[t]{1in}{\centering No. of Subsets / \par Length Distribution}\\
\hline\hline
l=0 & \{$\phi$\} & 0 & 1 \\ \hline
l=1 & \{1\} & 1 & 1 \\
\hline
\end{tabular}
\hfill
\begin{center}
\begin{tabular}[t]{|c|c|c|c|}
\hline
\parbox[t]{0.5in}{\centering Values of $l$ \par for $n=2$ } & Subset & Sum of the Subset & \parbox[t]{1in}{\centering No. of Subsets / \par Length Distribution}\\
\hline\hline
l=0 & \{$\phi$\} & 0 & 1 \\ \hline
\multirow{2}{*}{l=1} & \{1\} & 1 & 1 \\
& \{2\} & 2 & 1 \\ \hline
l=2 & \{1, 2\} & 3 & 1 \\ \hline
\end{tabular}
\end{center}
\caption{Length-Sum Distribution for base cases: $X_{0}, X_{1} $ and $X_{2}$. First column presents the possible length values, second and third column presents the corresponding subsets and their sum respectively and the fourth column presents the Length-Sum distribution, $LD[n][S][l]$. \label{Ch2TableLengthSumDisBaseCases}}
\end{table}

\begin{table}
\footnotesize
\begin{center}
\begin{tabular}[t]{|c|c|c|c|c|c|c|c|c|c|c|c|c|c|c|}
\hline
\multicolumn{15} {|c|} {Values for n=5} \\ \hline
\multicolumn{3} {|c|} {l=0}  & \multicolumn{3} {c|} {l=1} & \multicolumn{3} {c|} {l=2}  & \multicolumn{3} {c|} {l=3} & \multicolumn{3} {c|} {l=4} \\ \hline
Sum & Subset & \textbf{Size} & Sum & Subset & \textbf{Size} & Sum & Subset & \textbf{Size} & Sum & Subset & \textbf{Size} & Sum & Subset & \textbf{Size}\\ \hline \hline
\parbox[t]{0.1in}{\centering
 0} & \parbox[t]{0.1in}{\centering
 $\phi$} & \parbox[t]{0.1in}{\centering
 1} & 1 & \{1\} & 1 & 3 & \{1, 2\} & 1 & 6 & \{1, 2, 3\}  & 1 & 10 &  \{1, 2, 3, 4\} & 1\\
 & & &  2 & \{2\} & 1 & 4 & \{1, 3\} & 1  & 7 & \{1, 2, 4\} & 1 & 11 & \{1, 2, 3, 5\} & 1\\
 & & &  3 & \{3\} & 1 & 5 & \parbox[t]{0.4in}{\centering
\{1, 4\}, \par \{2, 3\}} & 2  & 8 & \parbox[t]{0.5in}{\centering
 \{1, 2, 5\}  \par  \{1, 3, 4\}}  & 2 & 12 & \{1, 2, 4, 5\} & 1\\
 & & &  4 & \{4\} & 1 & 6 & \parbox[t]{0.4in}{\centering
 \{1, 5\} \par \{2, 4\}} & 2  & 9 & \parbox[t]{0.5in}{\centering
 \{2, 3, 4\} \par  \{1, 3, 5\}} & 2 & 13 & \{1, 3, 4, 5\}  & 1\\
 & & &  5 & \{5\} & 1 & 7 &\parbox[t]{0.4in}{\centering
  \{2, 5\} \par  \{3, 4\}} & 2 & 10 & \parbox[t]{0.5in}{\centering
 \{2, 3, 5\} \par \{1, 4, 5\}} & 2 & 14 &  \{2, 3, 4, 5\}  & 1\\
 & & &  & & & 8 & \{3, 5\} & 1  & 11 & \{2, 4, 5\} & 1 & & & \\
 & & &  & & & 9 & \{3, 6\} & 1  & 12 & \{3, 4, 5\} & 1 & & & \\
\hline
\hline
\multicolumn{15} {|c|} {l=5}\\ \hline
\multicolumn{5} {|c|} {Sum} & \multicolumn{5} {c|} {Subset} & \multicolumn{5} {c|} {\textbf{Size}} \\ \hline\hline
\multicolumn{5} {|c|} {15} & \multicolumn{5} {c|} {\{1, 2, 3, 4, 5\}} & \multicolumn{5} {c|} {1} \\
\hline
\end{tabular}
\end{center}
\caption{Length-Sum Distribution for $\mathcal P \left({X_{5}}\right)$ \label{Ch2TableLengthSumDis5}}
\end{table}

\subsection{Element Distribution}
\label{Ch3ElementDistribution}
In Section \ref{Ch2:SumDistribution}, we have explained and explored the concept of Sum Distribution, where we count the number of subsets out of all power set $\mathcal P \left({X_{n}}\right) $, of $X_{n}$ which add up to a certain number $S$. Let us assume, $M$ represents such sets. We study the occurrence of each element from set $X_{n}$ in set $M$. $e$ denotes each element of $X_{n}$, $\forall e \in [1, n]$, $\forall S \in [0, \frac{n(n+1)}{2}]$, element distribution function, $ED[n][S][e]$, is defined as follows:
\begin{equation}  
\label{Ch3Eq1}
    ED[n][S][e] = \begin{cases}
               0 & (n=0) \  \text{or} \  (S=0) \  \text{or} \  (e=0)\\
                & \text{or} \ (0 < S < n \  \text{and} \  e == n) \\
               ED[n-1][S][e] & 0 \leq S < n \  \text{and} \  1 \leq e < n\\
               ED[n-1][S][e] + ED[n-1][S-n][e] & n \leq S \leq \frac{n(n-1)}{2} \  \text{and} \  1 \leq e < n  \  \text{and} \  n > 2 \\
               SD[n-1][S-n]  & n \leq S \leq \frac{n(n+1)}{2} \  \text{and} \  e == n \\
               SD[n][S] - ED[n][maxSum-S][e] & \frac{n(n-1)}{2} + 1 \leq S \leq \frac{n(n+1)}{2} \  \text{and} \  1 \leq e < n\\
               0 & \text{otherwise}
     \end{cases}
\end{equation}

Element distribution is another prepossessing procedure required for presenting various alternate enumeration techniques especially bucket algorithms introduced in Section \ref{Ch5EnumerationTechnique1}.

Table \ref{Ch3TableEleDisBaseCase} represents the count of elements in $\{1, 2\}$ and $\{1, 2, 3\}$ in all subsets of $X_{2}$ and $X_{3}$ respectively which are divided based on their sums. These are the base cases. Similarly, Table \ref{Ch3TableEleDisX5} represents distribution of elements of $X_{5}$ in $\mathcal P \left({X_{5}}\right)$, where subsets are categorized on the basis of their Sum. Element distributions of $X_{0}$ includes the count of element $0$ in subset $\phi$ with $Sum = 0$ . We assume $ED[0][0][0] = 0$. For a given $n$, the count of element $0$ in all the subsets is considered as $NULL$ or $0$. We are not including $0$ in the set of first $n$ natural numbers. This generate $ED[n][S][0] = 0 \  \forall S \in [0, \frac{n(n+1)}{2}]$. Also,  for any value of $n$, a zero-sum is achieved only by subset $\phi$ which is an empty set. Therefore,  $ED[n][0][e] = 0 \  \forall e \in [0, n]$. We consider values of elements distribution for $\mathcal P \left({X_{0}}\right)$, $\mathcal P \left({X_{1}}\right)$ and $\mathcal P \left({X_{2}}\right)$ as seed values. Following are the values:
\begin{enumerate}
\item $ED[0][0][0] = 0$
\item $ED[1][1][1] = ED[2][1][1] = 1$
\item $ED[2][2][2] = 1$
\item $ED[2][3][1] = ED[2][3][2] = 1$
\item otherwise $ED[i][j][k] = 0$ 
\end{enumerate}

\begin{table}[!htbp]
\footnotesize
\begin{tabular}{|c|c|c|c|c|}
\hline
Subsets $\rightarrow$ & $\phi$ & \{1\} & \{2\} & \{1, 2\} \\ \hline
Elements $\downarrow$ & \multicolumn{4} {l|} {} \\  \hline  \hline
1 & 0 & 1 & 0 & 1\\
2 & 0 & 0 & 1 & 1\\
\hline
\end{tabular}
\hfill
\hspace{3em}
\begin{tabular}{|c|c|c|c|c|c|c|c|c|}
\hline
Subsets $\rightarrow$ & $\phi$ & \{1\} & \{2\} & \{3\} & \{1, 2\} & \{1, 3\} & \{2, 3\} & \{1, 2, 3\} \\ \hline
Elements $\downarrow$ & \multicolumn{8} {l|} {} \\  \hline  \hline
1 & 0 & 1 & 0 & 0 & 1 & 1 & 0 & 1\\
2 & 0 & 0 & 1 & 0 & 1 & 0 & 1 & 1\\
3 & 0 & 0 & 0 & 1 & 0 & 1 & 1 & 1\\
\hline
\end{tabular}
\caption{Distribution of elements [1,2] in $\mathcal P \left({X_{2}}\right) $ and elements [1,2,3] in $\mathcal P \left({X_{3}}\right)$. \label{Ch3TableEleDisBaseCase}}
\end{table}

\begin{table}[!htbp]
\centering
\footnotesize
\begin{tabular}{|c|c|c|c|c|c|c|c|c|c|c|c|c|c|c|c|c|}
\hline
\multicolumn{17} {|c|} {Values for n=5} \\ \hline
\parbox[t]{1in}{\centering No. of Subsets \par for a Sum} & 1 & 1 & 1 & 2 & 2 & 3 & 3 & 3 & 3 & 3 & 3 & 2 & 2 & 1 & 1 & 1 \\ \hline
Sum $\rightarrow$ & 0 & 1 & 2 & 3 & 4 & 5 & 6 & 7 & 8 & 9 & 10 & 11 & 12 & 13 & 14 & 15\\ \hline
Integers $\downarrow$ & \multicolumn{16} {l|} {}  \\ \hline  \hline
1& 0 & 1 & 0 & 1 & 1 & 1 & 2 & 1 & 2 & 1 & 2 & 1 & 1 & 1 & 0 & 1 \\
2& 0 & 0 & 1 & 1 & 0 & 1 & 2 & 2 & 1 & 1 & 2 & 2 & 1 & 0 & 1 & 1 \\
3& 0 & 0 & 0 & 1 & 1 & 1 & 1 & 1 & 2 & 2 & 2 & 1 & 1 & 1 & 1 & 1\\
4& 0 & 0 & 0 & 0 & 1 & 1 & 1 & 2 & 1 & 2 & 2 & 1 & 2 & 1 & 1 & 1\\
5& 0 & 0 & 0 & 0 & 0 & 1 & 1 & 1 & 2 & 2 & 2 & 2 & 2 & 1 & 1 & 1\\
\hline
\end{tabular}
\caption{Distribution of elements $[1, 2, 3, 4, 5]$ in $\mathcal P \left({X_{5}}\right)$. \label{Ch3TableEleDisX5}} 
\end{table}

\subsubsection{Correctness of the Element Distribution Formula}
\label{Ch3:Correctness of the FormulaEle}
We present the theorems and lemma which prove the correctness of Element distribution formula, $ED[n][S][e]$ presented in Equation \ref{Ch3Eq1}. $ED[n][S][e]$ represents the count of element $e$ in those subsets of $X_{n}$ which has sum $S$ where $e \in [1, n]$, $S \in [0, maxSum]$ and $maxSum = \frac{n(n+1)}{2}$.
\begin{thm}
\label{Ch3ThmEleDis1}
$ED[n][S][n] =  0 $ if $0 < S < n$.
\end{thm}
\begin{proof}
Let us assume $ED[n][S][e] \neq 0$ and $ED[n][S][e] = c$, where $c$ is a positive integer. $c$ is the count of number of times an element $e$ occur in a class of subsets $element_{(n,S,e)}$ where $element_{(n,S,e)}$ consist of all the subsets of $\mathcal P \left({X_{n}}\right) $ which add up to a sum of $S$. Since, $c$ represents a count, it cannot be negative. By definition $c$, $ED[n][S][e]$ and $element_{(n,S,e)}$ follow these equations:
\begin{equation}
\label{Ch3Eq2}
c = |element_{(n,S,e)}|
\end{equation}   
\begin{equation}
\label{Ch3Eq3}
ED[n][S][e] = |element_{(n,S,e)}|
\end{equation}   
\begin{equation}
\label{Ch3Eq4}
c = ED[n][S][e]
\end{equation} 
\\
Let $A$ be a subset of $element_{(n,S,e)}$. Then, $e$ will belong to $A$ and sum of all elements of $A$ will be greater than or equal to $e$.
\begin{equation}
\label{Ch3Eq5}
e \in A
\end{equation}   
\begin{equation}
\label{Ch3Eq6}
Sum(A) \geq e
\end{equation}   
\begin{equation}
\label{Ch3Eq7}
S \geq e
\end{equation}   
Since $(e == n)$ as per the initial conditions, Equation \ref{Ch3Eq7} will become,
\begin{equation}
\label{Ch3Eq8}
S \geq n
\end{equation}
Since $ 0 \leq S < n$ it results into a contradiction. Our assumption is false. There are no subsets which contain $e$ and have sum less than $e$. Therefore, from the condition $c = 0$ and from Equation \ref{Ch3Eq7}
\begin{equation}
\label{Ch3Eq9}
ED[n][S][e] = 0
\end{equation} 
\begin{equation}
\label{Ch3Eq9-1}
ED[n][S][e] =  0 \  if \  0 < S < n \  and \  e == n
\end{equation} 
Hence, we have proved the first part of Equation \ref{Ch3Eq1}.
\end{proof}

\begin{thm}
\label{Ch3ThmEleDis2}
$ED[n][S][e] = ED[n-1][S][e] + ED[n-1][S-n][e]$ if $n \leq S \leq \frac{n(n-1)}{2}$, $1 \leq e < n$ and $n > 2$.
\end{thm}
\begin{proof}
Let $element_{(n,S,e)}$ be a class of subsets which consists of all the subsets of $P(X_{n})$ which sum upto $S$ and contain an element $e$, $1 \leq e < n$. Let us assume, a set $A \in element_{(n,S,e)}$. Since $(S \geq n)$, then $A$ may or may not contain element $n$. If $n \in A$ then $A-n$ belongs to the class of subsets of $\mathcal P \left({X_{n-1}}\right)$ which sum upto $(S-n)$ and contain an element $e$ (as presented in Equation \ref{Ch3Eq10}). If $n \notin A$ then, $A$ belongs to the class of subsets of $\mathcal P \left({X_{n-1}}\right)$ which sum upto $S$ and contain an element $e$ (as presented in Equation \ref{Ch3Eq11}).
\begin{equation}
\label{Ch3Eq10}
A-n \in element_{(n-1,S-n,e)}
\end{equation}
\begin{equation}
\label{Ch3Eq11}
A \in element_{(n-1,S,e)}
\end{equation}
From Equation \ref{Ch3Eq10} and Equation \ref{Ch3Eq11}, we form the set of all subsets which sum up to $S$ and contain element $e$,
\begin{equation}
\label{Ch3Eq12}
element_{(n,S,e)} = element_{(n-1,S,e)} \cup  element_{(n-1,S-n,e)} \quad n \leq S \leq \frac{n(n-1)}{2} \quad and \quad 1 \leq e < n
\end{equation}
Taking cardinality on both sides of Equation \ref{Ch3Eq12},
\begin{equation}
\label{Ch3Eq13}
|element_{(n,S,e)}| = |element_{(n-1,S,e)}| + |element_{(n-1,S-n,e)}| \quad n \leq S \leq \frac{n(n-1)}{2} \quad and \quad 1 \leq e < n
\end{equation}
\begin{equation}
\label{Ch3Eq14}
ED[n][S][e] = ED[n-1][S][e] + ED[n-1][S-n][e] \quad n \leq S \leq \frac{n(n-1)}{2} \quad and \quad 1 \leq e < n
\end{equation}

In order to complete this proof following properties of $element_{(n,S,e)}$ should be proved.
\begin{enumerate}
\item \textit{Uniqueness:} There should be no duplicate subsets in $element_{(n,S,e)}$, $element_{(n-1,S,e)} \cap element_{(n-1,S-n,e)} = \phi$.

\begin{proof}
$element_{(n-1,S,e)}$ is the set of all the subsets of $\mathcal P \left({X_{(n-1)}}\right)$ containing element $e$ with sum $S$ and $element_{(n-1,S-n,e)}$ is the set of all the subsets of $\mathcal P \left({X_{(n-1)}}\right)$ containing element $e$ with sum $(S-n)$. We use the method of contradiction to prove set of subsets in $element_{(n-1,S,e)}$ and $element_{(n-1,S-n,e)}$ are independent. Let us assume, subset $p$ belongs to both $element_{(n-1,S,e)}$ and $element_{(n-1,S-n,e)}$. Since, $p \in element_{(n-1,S,e)}$, therefore by definition, the subset $p$ contains element $e$, has elements ranging from $1$ to $(n-1)$ and these elements sum upto $S$.
\begin{equation}
\label{Ch3Eq59}
S = \sum_{i=1}^{len(p)} p_{i}
\end{equation}

Similarly, as per assumption, $p \in element_{(n-1,S-n,e)}$. Therefore by definition, the subset $p$ contains element $e$, has elements ranging from $1$ to $(n-1)$ and these elements sum upto $(S-n)$.
\begin{equation}
\label{Ch3Eq60}
(S-n) = \sum_{i=1}^{len(p)} p_{i}
\end{equation}

From Equation \ref{Ch3Eq59} and Equation \ref{Ch3Eq60}, there is a contradiction as $\sum_{i=1}^{len(p)} p_{i}$ is both $S$ and $(S-n)$. Since, $n$ is a natural number, the above equations contradict our assumption that a subset $p$ can belong to both sets $element_{(n-1,S,e)}$ and $element_{(n-1,S-n,e)}$. Therefore, by contradiction, there is no subsets $p$ which belongs to both sets. Hence, $element_{(n-1,S,e)}$ and $element_{(n-1,S-n,e)}$ are independent.
\end{proof}

\item \textit{Completeness:} $element_{(n,S,e)}$ should contain all the subsets of $\mathcal P \left({X_{n}}\right)$ which contain element $e$ and sum upto $S$.
\begin{proof}
The power set of $X_{n}$, $\mathcal P \left({X_{n}}\right)$ which contain element $e$ and sum upto $S$ can be divided into two parts: subsets with sum $S$ which contain element $n$ and subsets with sum $S$ which do not contain element $n$. By definition, $element_{(n-1,S,e)}$ is the set of all the subsets of $\mathcal P \left({X_{(n-1)}}\right)$ with sum $S$ containing element $e$ and $element_{(n-1,S-n,l-1)}$ is the set of all the subsets of $\mathcal P \left({X_{(n-1)}}\right)$ with sum $(S-n)$ containing element $e$.

In Equation \ref{Ch3Eq12}, the union of sets $element_{(n-1,S,e)}$ and $element_{(n-1,S-n,e)}$ generates all subsets of $\mathcal P \left({X_{n}}\right)$ with sum $S$ containing element $e$. Therefore, $element_{(n,S,e)}$ should consists of subsets of $\mathcal P \left({X_{n}}\right)$ with sum $S$ containing element $e$.
\end{proof}
\end{enumerate}

The above two proofs are required to complete the statement: $ED[n][S][e] = ED[n-1][S][e] + ED[n-1][S-n][e]$ if $n \leq S \leq \frac{n(n-1)}{2}$ and $1 \leq e < n$. This theorem will only be true, if sum is positive i.e. $S \geq 0$
\begin{equation}
\label{Ch3Eq15}
S \geq 0
\end{equation}
\begin{equation}
\label{Ch3Eq16}
\frac{n(n-1)}{2} - n \geq 0
\end{equation}
\begin{equation}
\label{Ch3Eq17}
\frac{n^2-n-2n}{2} \geq 0
\end{equation}
\begin{equation}
\label{Ch3Eq18}
\frac{n^2-3n}{2} \geq 0
\end{equation}
\begin{equation}
\label{Ch3Eq19}
\frac{n(n-3)}{2} \geq 0
\end{equation}
\begin{equation}
\label{Ch3Eq20}
n(n-3) \geq 0
\end{equation}
Therefore, either both $n$ and $n-3$ should be greater than $0$ or both should be less than $0$. Since, $n$ cannot be negative,
\begin{equation}
\label{Ch3Eq21}
n \geq 0 \quad and \quad n \geq 3
\end{equation}
Therefore,
\begin{equation}
\label{Ch3Eq22}
n \geq 3
\end{equation}

Hence, from Equation \ref{Ch3Eq14} and Equation \ref{Ch3Eq22} we have proved the third part of Equation \ref{Ch3Eq1}.
\end{proof}

\begin{lem}
$ED[n][S][e] = ED[n-1][S][e]$ if $0 \leq S < n$ and $1 \leq e < n$.
\end{lem}
\begin{proof}
According to Theorem \ref{Ch3ThmEleDis2}, 
\begin{equation}
\label{Ch3Eq23}
ED[n][S][e] = ED[n-1][S][e] + ED[n-1][S-n][e] \quad n \leq S \leq \frac{n(n-1)}{2} \quad and \quad 1 \leq e < n
\end{equation}
Since,
\begin{equation}
\label{Ch3Eq24}
0 \leq S < n
\end{equation}
\begin{equation}
\label{Ch3Eq25}
(-n) \leq S-n < 0
\end{equation}
But a sum cannot be negative. Therefore, count of element $e$ in subsets of $\mathcal P \left({X_{n-1}}\right) $ which sum up to $S$ is zero, $ED[n-1][S-n][e] = 0$.
\begin{equation}
\label{Ch3Eq26}
ED[n][S][e] = ED[n-1][S][e] + 0
\end{equation}
\begin{equation}
\label{Ch3Eq27}
ED[n][S][e] = ED[n-1][S][e]  \quad  0 \leq S < n  \quad  and  \quad  1 \leq e < n
\end{equation}
Equation \ref{Ch3Eq27} proves the second part of Equation \ref{Ch3Eq1}. 
\end{proof}

\begin{thm}
\label{Ch3ThmEleDis3}
$ED[n][S][e] = SD[n-1][S-n]$ if $n \leq S \leq \frac{n(n+1)}{2}$ and $e == n$.
\end{thm}
\begin{proof}
Let $element_{(n,S,e)}$ be a class of subsets where it consist of all the subsets of $\mathcal P \left({X_{n}}\right)$ which sum up to $S$ and contain an element $e$, $e==n$. Let us assume $ A \in element_{(n,S,e)}$ and $|element_{(n,S,e)}| = ED[n][S][e] = c$ where $c \geq 0$.  Since, element $e$ belongs to set $A$, $e \in A$,
\begin{equation}
\label{Ch3Eq28}
A-e \equiv A-n \in element_{(n-1,S-e,0)}
\end{equation}
Sum $S$ will result in following condition,
\begin{equation}
\label{Ch3Eq29}
n \leq S \leq \frac{n(n+1)}{2}
\end{equation}
\begin{equation}
\label{Ch3Eq30}
0 \leq S-n \leq \frac{n(n+1)}{2} - n
\end{equation}
Let us assume $S-n$ as $S'$,
\begin{equation}
\label{Ch3Eq31}
0 \leq S' \leq \frac{n^2-n}{2}
\end{equation}
\begin{equation}
\label{Ch3Eq32}
0 \leq S' \leq \frac{n(n-1)}{2}
\end{equation}
\begin{equation}
\label{Ch3Eq33}
maxSum(n-1) = \frac{n(n-1)}{2}
\end{equation}
From Equation \ref{Ch3Eq28} and Equation \ref{Ch3Eq33},
\begin{equation}
\label{Ch3Eq34}
\forall A-{n} \in element_{(n-1,S-n,0)} \equiv element_{(n-1,S-n,e')} \quad where \quad 1 \leq e' < n 
\end{equation}
\begin{equation}
\label{Ch3Eq35}
\forall A \in element_{(n,S-n+n,n)} \equiv element_{(n-1,S-n,e')} \quad where \quad 1 \leq e' < n 
\end{equation}
\begin{equation}
\label{Ch3Eq36}
\forall A \in element_{(n,S,n)} \equiv element_{(n-1,S-n,e')} \quad where \quad 1 \leq e' < n 
\end{equation}
Taking cardinality on both sides,
\begin{equation}
\label{Ch3Eq37}
|element_{(n,S,n)}|  = |element_{(n,S,r)}| = |element_{(n-1,S-n,e')}|
\end{equation}
\begin{equation}
\label{Ch3Eq38}
ED[n][S][e==n] = ED[n-1][S-n][e']
\end{equation}

\noindent $|element_{(n-1,S-n,e')}|$ is the number of subsets $X_{n-1}$ which sum up to $(S-n) = (S-e) = (S-n)$ . By using the concept of sum distribution defined in Section \ref{Ch2:SumDistribution} and Equation \ref{Ch3Eq38},
\begin{equation}
\label{Ch3Eq39}
|element_{(n-1,S',e')}| = SD[n-1][S-n]
\end{equation}
\begin{equation}
\label{Ch3Eq40}
ED[n-1][S'][e'] = ED[n][S][e==n] = SD[n-1][S-n]
\end{equation}
\begin{equation}
\label{Ch3Eq41}
ED[n][S][e] = SD[n-1][S-n] \quad where \quad n \leq S \leq \frac{n(n+1)}{2} \quad and \quad e == n
\end{equation}
Equation \ref{Ch3Eq41} proves the fourth part of Equation \ref{Ch3Eq1}. 
\end{proof}

\begin{thm}
\label{Ch3ThmEleDis4}
$ED[n][S][e] = SD[n][S] - ED[n][maxSum(n)-S][e]$ if  $(\frac{n(n-1)}{2} + 1) \leq S \leq \frac{n(n+1)}{2}$ and $1 \leq e < n$
\end{thm}

\begin{proof}
$maxSum(n)$ is the sum of all elements of $X_{n} = 1 + 2 + \ldots n = \frac{n(n+1)}{2}$, as defined in Section \ref{Ch2sec:Formulation}. Let us assume $S' = maxSum(n) - S$. Since, the plot between number of subsets and sum follow a Gaussian symmetric distribution, $SD[n][S]$ will be equal to  $SD[n][maxSum(n) - S]$.
\begin{equation}
\label{Ch3Eq42}
SD[n][S] = SD[n][S'] = c
\end{equation}
There are $c$ number of subsets which sum up to $S$ and $S'$.  In this case, sum $S$ is greater than the $\frac{maxSum}{2}$ (the mid point) and by using the reflection/symmetric property of the curve we can find all the values of $ED[n][S][e]$.
\begin{equation}
\label{Ch3Eq43}
S'' = \frac{maxSum}{2} = \frac{n(n+1)}{4}
\end{equation}
\begin{equation}
\label{Ch3Eq44}
S_{low} = \frac{n(n-1)}{2} + 1
\end{equation}
\begin{equation}
\label{Ch3Eq45}
S_{low} - S'' = \frac{n(n-1)}{2} + 1 - \frac{n(n+1)}{4}
\end{equation}
\begin{equation}
\label{Ch3Eq46}
S_{low} - S'' = \frac{n(2n-2-n+1)}{2} + 1
\end{equation}
\begin{equation}
\label{Ch3Eq47}
f(n) = S_{low} - S'' = \frac{n^2-3n+2}{2}
\end{equation}
By using the property of second derivative test we show that $f'(n)$ is greater than $0$ when $S > \frac{maxSum}{2}$.
\begin{equation}
\label{Ch3Eq48}
f'(n) = \frac{d(f(n))}{dn} > 0 
\end{equation}
\begin{equation}
\label{Ch3Eq49}
f'(n) = \frac{d(n^2-3n+2/2)}{dn} > 0 
\end{equation}
\begin{equation}
\label{Ch3Eq50}
f'(n) = n - \frac{3}{2} > 0 
\end{equation}
\begin{equation}
\label{Ch3Eq51}
f'(n) = n > \frac{3}{2}
\end{equation}
Therefore, $\forall n \geq 2$ we can use the symmetric property and calculate half of the values by using the previously calculated values. For $n=1$ values of element distribution will be covered as the part of base cases. 
\\
\noindent Let $sum_{(n,S)}$ be a set of all the subsets of $\mathcal P \left({X_{n}}\right) $ which sum up to $S$ and $sum_{(n,S')}$ consist of all subsets of $\mathcal P \left({X_{n}}\right) $ which sum to $S'$, where $S' = (maxSum - S)$. $\forall A \in sum_{(n,S)}$ and $A^c \in sum_{(n,S')}$ where $A^c$ is the complement set of $A$.
\begin{equation}
\label{Ch3Eq52}
A \cup A^c = U
\end{equation}
Since, $U$ is the universal set, $U = \{1, 2 \ldots n\}$ and contain a single occurrence of each element $e \in [1, n]$, therefore, $A \cup A^c$ also contains a single occurrence of each element $e$. From Equation \ref{Ch3Eq52} there are $c$ subsets in $A$ and $A^c$. $\forall k \in [1, n]$ count of $e$ in $A$ and $A^c$ is $1$. Let us define $Count(x,y)$ as the count of element $x$ in any subset or class of subsets $y$.
\begin{equation}
\label{Ch3Eq53}
Count(e, A) + Count(e, A^c) = 1
\end{equation}
$\forall e \in [1, n], \forall A \in sum_{(n,S)} \  and \  \forall A^c \in sum_{(n,S')} $
\begin{equation}
\label{Ch3Eq54}
Count(e,  sum_{(n,S)}) + Count(e,  sum_{(n,S')}) = |sum_{(n,S)}| * 1 = |sum_{(n,S')}| * 1
\end{equation}
By using the definition of element distribution and Equation \ref{Ch3Eq42}
\begin{equation}
\label{Ch3Eq55}
ED[n][S][e] + ED[n][S'][e] = c
\end{equation}
\begin{equation}
\label{Ch3Eq56}
ED[n][S][e] + ED[n][S'][e] = SD[n][S]
\end{equation}
\begin{equation}
\label{Ch3Eq57}
ED[n][S][e] = SD[n][S] - ED[n][S'][e]
\end{equation}
Therefore, by putting the value of $S' = (maxSum(n) - S)$
\begin{equation}
\label{Ch3Eq58}
ED[n][S][e] = SD[n][S] - ED[n][maxSum(n) - S][e]
\end{equation}
Equation \ref{Ch3Eq58} proves the last part of Equation \ref{Ch3Eq1}. 
\end{proof}

Sum Distribution, Length-Sum Distribution and Element Distribution are used in developing alternate enumeration techniques for solving SSP. These techniques are presented in the next section.

\section{Alternate Enumeration Techniques for Subset Sum Problem}
\label{AlternateEnuTechq}
In this paper, we propose seven approaches to find the solution for enumerating all the $(2^{n}-1$) subsets of $X_{n}$. In each approach, we choose different method for addressing the enumeration of SSP. We propose novel algorithms: Subset Generation using Sum Distribution (SDG), Subset Generation using Length-Sum Distribution (LDG), Basic Bucket Algorithm (Basic BA), Maximum and Minimum Frequency Driven Bucket Algorithms (Max FD and Min FD) and Local Search using Maximal and Minimal Subsets (LS MaxS and LS MinS) for enumerating SSP. The first approach is the backtracking algorithm. It is the naive method for solving SSP. This algorithm is used to benchmark the new proposed algorithms. 

\subsection{Subset Generation using Backtracking}
\label{Ch1Backtracking}
Our aim is to find all the subsets of set $X_{n}$ with $Sum=S$. According to the exhaustive search algorithm for SSP \cite{Authors03Wiki}, we try to find the resulting subset by iterating through all possible $2^{n}$ solutions. But in this algorithm, we arrange the elements in an orderly fashion. The worst case time complexity for this algorithm is exponential. It is $\mathcal{O}(n \times 2^{n})$. The space complexity for this algorithm is the size of the input, $\mathcal{O}(n)$. Even though backtracking is a clean and crisp algorithm for SSP, this algorithm has many drawbacks. It tries to generate all the desired subsets by checking every branch and subset. Since there can be a lot of high branches at every state of the back tracking algorithm, this leads to inefficient, multiple recursive calls and reversion to old states. It requires a large amount of time and space to reflect the changes in the system stack.

\subsection{Subset Generation using Sum Distribution}
\label{Ch5SumDisGen}
We design a generator using Sum Distribution. Algorithm \ref{Ch4SumDistributionGen} is the pseudo-code for generating all the subsets of $X_{n}$ with sum $S$. As we know, sum distribution is recursive and uses subsets of $X_{(n-1)}$ to produce results for $X_{n}$. We store these previous values with the help of $SDG$ (initialized at Line $1$). Extra values of $SDG$ ($SDG[i-1]$) are freed in Line $20$ to minimize the space consumption. In Line $2$, we iterate through smaller natural numbers. Line $3$ to Line $6$ define $start\_sum$, $mid\_sum$, $end\_sum$ and $universal\_set$. Line $7$ to Line $19$ iterate through values of sum between $start\_sum$ and $mid\_sum$. The desired set of subsets, $SDG[i][j]$ (subsets of $X_{i}$ with sum $j$), consists of all subsets of $SDG[i-1][j]$ and $SDG[i-1][j-i]$. Next, we include $i^{th}$ element in every subset of $SDG[i-1][j-i]$. For each of these resulting subsets, a symmetric subset of sum $(end\_sum - j)$ is calculated by subtracting the subset from $universal\_set$. Line $11$ to Line $18$ essentially execute these steps and returns the final result at Line $22$.

The value of maximum number of subsets has exponential bound, $\mathcal{O}(2^n * n^{\frac{-3}{2}})$, as described in Appendix \ref{Chp5SumDistribution}. Therefore, the time complexity for $(loop_{3})$ at Line $14$ is $\mathcal{O}(2^n * n^{\frac{-3}{2}})$. Since, $n \in [1, n]$ and $S \in [0,\frac{n(n+1)}{2}]$, time complexity of the above algorithm results to $ \mathcal{O}(loop_{1})* \mathcal{O}(loop_{2}) * \mathcal{O}(loop_{3}) = \mathcal{O}(n) * \mathcal{O}(n^2) * \mathcal{O}(2^n * n^{\frac{-3}{2}}) = \mathcal{O}(2^n * n^{\frac{3}{2}})$. Space complexity for the above algorithm is the size of array storing smaller subsets, $SDG[n-1][S]$. This complexity is also exponential $n * S * \ No. \ of \ Subsets$. Since $S \in [0, \frac{n(n+1)}{2}]$, the space complexity results to $ \mathcal{O}(n)* \mathcal{O}(n^2) * \mathcal{O}(2^n * n^{\frac{-3}{2}})$ i.e. $\mathcal{O}(2^n * n^{\frac{3}{2}})$.

\begin{algorithm}
  \caption{SDG: GeneratorUsingSumDistribution($n$)}
  \label{Ch4SumDistributionGen}
  \begin{algorithmic}[1]
  	\State $SDG = \{{}\} $\Comment{Data structure to store the generated Subsets}
    \For{$i \in \{1,\dots,n\}$}
      \State $start\_sum = 0$
	  \State $mid\_sum = \floor{\frac{i(i+1)}{4}}$
      \State $end\_sum = \frac{i(i+1)}{2}$ \Comment{$end\_sum$ is equal to $maxSum(i)$}
      \State $universal\_set = \{1, 2 \ldots n\}$ \Comment{$universal\_set$ is used to calculate the symmetric subsets}
      \For{$j \in \{start\_sum,\dots,mid\_sum\}$}
      	\If {$(j == 0)$}
      		\State $SDG[i] = \{\phi\}$
      	\EndIf
	    \State $SDG[i][j] = SDG[i-1][j]$
       	\For{$subset \in SDG[i-1][j-i]$}
       		\State $subset.append(i)$
	    	\State $SDG[i][j].append(subset)$ \Comment{Adding $i^{th}$ element in every subset of $SDG[i-1][j-i]$}
	    \EndFor
        \If{$j \neq (i-j)$}
		    \State $SDG[i][end\_sum-j] = universal\_set - SDG[i][j]$ \Comment{Symmetric subsets.}
	    \EndIf
      \EndFor
      \State $Free(SDG[i-1])$
    \EndFor
    \State \Return $SDG[n]$
  \end{algorithmic}
\end{algorithm}

\subsection{Subset Generation using Length-Sum Distribution}
\label{Ch5LengthSumDisGen}
Along with Sum Distribution, we have established several concepts, theories and formulas for $Length-Sum$ Distribution as well. It counts the number of subsets of $X_{n}$ of length $l$ and sum $S$ where $X_{n} = \{1,2,3 \ldots n\}$, $l \in [0, n]$ and $S \in [0, \frac{n(n+1)}{2}]$, represented by $LD[n][S][l]$. The recursive equation (Equation \ref{Ch2Eq12}) establishes the theory for the Length-Sum distribution.

In this section, we present the designed generator. Algorithm \ref{Ch4LenSumDistributionGen} is the pseudo-code for generating all the subsets of $X_{n}$ of length $l$ and sum $S$. This distribution is recursive and uses $LDG$ to store the previous output which is initialized at Line $1$ and Line $10$. The notation for $LDG$ is different than notation of $LD$. We denote the count the number of subsets of $X_{n}$ of length $l$ and sum $S$ where $X_{n} = \{1,2,3 \ldots n\}$, $l \in [0, n]$ and $S \in [0, \frac{n(n+1)}{2}]$ by $LD[n][S][l]$. However, $LDG[i][j][k]$ consists of all subsets of $X_{i}$ with $length = j$ and $Sum = k$. In $LDG$ notation for length and sum are reversed for easier calculations.

In Algorithm \ref{Ch4LenSumDistributionGen}, extra values of $LDG$ ($LDG[i-1]$) are freed in Line $26$ to minimize the space consumption. In Line $5$, Line $9$ and Line $13$, we iterate through smaller natural numbers, length range and possible values of sum respectively. Line $6$ to Line $12$ we define $max\_sum$ for $X_{i}$, bases cases of $LDG[i][j]$,  $start\_sum$ and $end\_sum$. Line $13$ to Line $24$ iterates through feasible values of sum between $start\_sum$ and $end\_sum$. The desired set of subsets, $LDG[i][j][k]$ consists of all subsets of $LDG[i-1][j][k]$ and $LDG[i-1][j-1][k-i]$. We include $i^{th}$ element in every subset of $LDG[i-1][j-1][k-i]$. For each of these resulting subsets, a symmetric subset of length $(i-j)$ and sum $(end\_sum - k)$ is calculated by subtracting the subset from $universal\_set$. Line $15$ to Line $23$ essentially execute these steps and returns the final result at Line $28$.

The value of maximum number of subsets has exponential bound,  $\mathcal{O}(2^n * n^{\frac{-3}{2}})$, as described in Appendix \ref{Chp5SumDistribution}. Therefore, the time complexity for $(loop_{4})$ in Line $16$ is $\mathcal{O}(2^n * n^{\frac{-3}{2}})$.  
Since, $l \in [1, n]$ and $S \in [0, \frac{n(n+1)}{2}]$ time complexity of the above algorithm results to $ \mathcal{O}(loop_{1})\ * \  \mathcal{O}(loop_{2}) \ * \ \mathcal{O}(loop_{3}) \ * \ \mathcal{O}(loop_{4})$ i.e. $\mathcal{O}(n) * \mathcal{O}(n) * \mathcal{O}(n^2) * \mathcal{O}(2^n * n^{\frac{-3}{2}}) = \mathcal{O}(n^4 * 2^n * n^{\frac{-3}{2}}) = \mathcal{O}(2^n n^{\frac{5}{2}})$. Space complexity for the above algorithm is the size of array storing smaller subsets, $LDG[n-1]$. This complexity is also exponential $n * l * S * \ (No. \ of \ Subsets)$. Since, $l \in [1,n]$ and $S \in [0, \frac{n(n+1)}{2}]$ the space complexity results to $ \mathcal{O}(n)* \mathcal{O}(n) * \mathcal{O}(n^2) * \mathcal{O}(2^n * n^{\frac{-3}{2}}) = \mathcal{O}(n^4)* \mathcal{O}(2^n * n^{\frac{-3}{2}})$ i.e. $\mathcal{O}(2^n * n^{\frac{5}{2}})$.

\begin{algorithm}
  \caption{LDG: GeneratorUsingLengthSumDistribution($n$)}
  \label{Ch4LenSumDistributionGen}
  \begin{algorithmic}[1]
  	\State $LDG=\{\}$
  	\State $LDG[0][0] = LD[1][0] = \{\}$ \Comment{Base Cases}
  	\State $LD[1][1] = \{[1]\}$ \Comment{Base Cases}
	\State $universal\_set = [1, 2 \ldots n]$ \Comment{$universal\_set$ is used to calculate the symmetric subsets}
	\For{$i \in \{2,\dots,n\}$}
	  \State $max\_sum = \frac{i(i+1)}{2}$
	  \State $LDG[i][0] = \{[1]\}$
	  \State $LDG[i][max\_sum] = \{universal\_set\}$
      \For{$j \in \{1,\dots,\frac{i}{2}\}$} \Comment{Iterarting till mid point}
		\State $LDG[i][j] = \{\}$
	    \State $start\_sum = \frac{j(j+1)}{2}$
        \State $end\_sum = i*j - \frac{i(i-1)}{2}$
        \For{$k \in \{start\_sum,\dots,end\_sum\}$}
          \State $LDG[i][j][k] = LDG[i-1][j][k]$
          \If{ $j \geq 1$ and $k \geq i$ and $i \leq k \leq \frac{i(i+1)}{2}$}
          	  \For{$subset \in LDG[i][j-1][k-i]$}
          	  	\State $subset.append(i)$
          	  	\State $LDG[i][j][k].append(subset)$ \Comment{Adding $i^{th}$ element in every subset of $LDG[i-1][j-1][k-i]$}
          	  \EndFor
          \EndIf
		  \If {$ j \neq (i-j)$}
		    \State $LDG[i][i-j][end\_sum-k] = universal\_set - LDG[i][j][k]$ \Comment{Symmetric subsets}
		  \EndIf
	    \EndFor
      \EndFor
      \State $Free(LDG[i-1])$
    \EndFor
    \State $Return \  LD[n]$
  \end{algorithmic}
\end{algorithm}

\subsection{Subset Generation using Basic Bucket Algorithm}
\label{Ch5EnumerationTechnique1}
In this section, we present a new method which generate all the subsets of $X_{n}$ with a particular sum. This is a greedy algorithm. The look-up table that has been used, has been explained in Section \ref{Ch5Lookup Table}. It has been extensively used with this algorithm.

The core idea behind this enumeration technique is to use the various distribution values that we have calculated so far, to construct all the subsets of $X_{n}$ which sum up to $S$. 

\noindent \textbf{Given:} The first concept used for Basic Bucket Algorithm is Element Distribution. We start with the exact occurrence of each element of $X_{n}$ in subsets of precise sum, $S$. This information is denoted by $ED[n][S][e]$. The next concept used is the number of subsets, among power set of $X_{n}$, where summation of all elements is $S$. $SD[n][S]$ denotes such count. For this algorithm, we consider $SD[n][S]$ as number of empty buckets. Buckets are storage data structures which are used to stack all the appropriate elements that compute the total sum $S$. We iterate through all elements in descending order. During each iteration, an element is assigned to one of the buckets. This method is about adding the correct element to the corresponding subset.
 
\noindent \textbf{Properties:} Element distribution and below properties help us ensure the correct placement for every element.
\begin{enumerate}
\item An element $e$ is added to a bucket $b$ only if the addition results to the uniqueness among all existing elements of the bucket $b$. This property is followed to guarantee that the generated result is a subset and it is not a bag. A subset belongs to power sets of $X_{n}$ $\mathcal P \left({X_{n}}\right)$.
\item An element $e$ is added to a bucket $b$ only if the addition of the element results to uniqueness amongst all the buckets. We follow this property to ensure the generation of correct number of subsets of sum $S$.
\item An element $e$ is added to a bucket $b$ only if on adding the new element, the sum of the bucket does not the exceed the desired sum $S$. This property allows us to create subsets of sum $S$.
\end{enumerate}

Unfortunately, we have no rule which forces only the generation of subsets with sum $S$. Many subsets with sum less than $S$ are generated during the first iteration of this technique. These subsets are called the $undesired$ set. For every subset of the $undesired$ set, $A$ $Sum(A)$ is less than $S$ i.e. $Sum(A) < S$. Therefore, we have converted this technique to a greedy algorithm. Instead of using this as a one time procedure, we reapply it with modified values of element distribution, $ED[n][S][e]$ and sum distribution, $SD[n][S]$. All subsets are generated by applying the same technique on modified input in a greedy manner. 

\noindent \textbf{Uniqueness:} The key step in successfully generating the full desired results is to maintain an efficient and complete lookup table as described in Section \ref{Ch5Lookup Table}. This lookup table which is maintained with the help of a hash function and bit vectors, not only ensures uniqueness among and within the buckets but also makes sure that all the $undesired$ subsets of previous iterations are properly hashed. So, we do not re-generate the same $undesired$ set in the next iteration. We need to put extra effort to preserve the state of all $undesired$ sets from every iteration. The whole lookup table is no bigger than $2^n$ and every subset: desired or undesired, is stored in the form of one integer $num$, where $num \in [0, 2^{n}]$. With the aim of preserving the count of every element from the set $X_{n}$, we maintain a log table for each round of iterations. The value of log table for each element, at the start of every round is the summation of value of element distribution at the end of last iteration of previous round and the count of all these elements from buckets which do not provide a subset of desired sum.

Algorithm \ref{Ch5AlgGetED} calculates the element distribution before start of each round of Basic Bucket Algorithm. Algorithm \ref{Ch5AlgInitializeBuckets} initializes the buckets at the start of the algorithm. It finds the value of $x$ and accordingly fill the buckets with the starting elements. This method is called from Line $3$ of the function \Call{GeneratingSubsets}{$n, S, SD[n][S], ED[n][S], prevWrongSubsets$} from the main Algorithm \ref{Ch5AlgGenerateSubsetsED}. We find an appropriate bucket for every element based on the properties of the Basic Bucket Algorithm. Functionality is defined in Algorithm \ref{Ch5AlgFindBucket}. While Algorithm \ref{Ch5AlgMainFunction} iterates though all the rounds of the bucket algorithm. All iterations of every round is implemented by the Algorithm \ref{Ch5AlgGenerateSubsetsED}.

\begin{algorithm}
  \caption{Basic BA: GetED($n , S, Table, wrongSubsets$)}
  \label{Ch5AlgGetED}
  \begin{algorithmic}[1]
	    \Function{GetED}{$n , S, Table, wrongSubsets$}
	    	\State $newTable = Table$
	    	\For {$subset \in wrongSubsets$}
	    		\For {$ele \in subset$}
	    			\State $newTable[ele] += 1$ \Comment{Restoring all the ellments of $wrongSubsets$ to the element distribution}
	    		\EndFor
	    	\EndFor 
	    	\State $Return \ newTable$
	    \EndFunction
    \end{algorithmic}
\end{algorithm}

\begin{algorithm}
  \caption{Basic BA: InitializeBuckets($all\_buckets$, $Table$, $n$, $S$ , $p$)}
  \label{Ch5AlgInitializeBuckets}
  \begin{algorithmic}[1]
    \Function{InitializeBuckets}{$all\_buckets$, $Table$, $n$, $S$ , $p$}
    	\State $q =$ count of non-zero entries of $Table$
    	\State $x = min(p, q)$
    	\State $elements = x$ largest integers of $X_{n}$ where $Table[ele] \neq 0$ $\forall \  ele \in elements$
    	\State Sort $elements$ in descending order
    	\State $bucketIndex = 1$
    	\For {$ele$ in $elements$}
    		\State Add $ele$ in $all\_buckets[bucketIndex]$
    		\State $bucketIndex++$
    	\EndFor
    \EndFunction
  \end{algorithmic}
\end{algorithm}

\begin{algorithm}
  \caption{Basic BA: FindBucket($all\_buckets$, $ele$, $S$)}
  \label{Ch5AlgFindBucket}
  \begin{algorithmic}[1]
	    \Function{FindBucket}{$all\_buckets$, $ele$, $S$}
	    	\For {$bucket$ in $all\_buckets$}
	    		\If {any $bucket$ entry is same as $ele$}
	    			\State Next
	    		\ElsIf {on adding $ele$ in $bucket$, $Sum(bucket) > S$}
	    			\State Next
	    		\ElsIf {on adding $ele$ in $bucket$, $bucket$ becomes duplicate to any other $bucket$}
	    			\State Next
	    		\Else
	    			\State \  Return$\  bucket$
	    		\EndIf
	    	\EndFor
  			\State $Return\  False$
	    \EndFunction
    \end{algorithmic}
\end{algorithm}

\begin{algorithm}
  \caption{Basic BA: Generating Subsets($n, S, SD[n][S], ED[n][S], prevWrongSubsets$)}
  \label{Ch5AlgGenerateSubsetsED}
  \begin{algorithmic}[1]
  	\State Given: $n$, $S$, $SD[n][S]$ and $ED[n][S][i]$ where $i \in [1, n]$ 
  	\State $desiredSubsets = [ \  ]$ \Comment{$desiredSubsets$ are all the subsets of $X_{n}$ with sum $S$.}
  	  	\State $wrongSubsets = [ \  ]$ \Comment{$wrongSubsets$ are the set of $undesired$ and $smaller $subsets.}
  	\State $Table = GetED(n, S, ED[n][S], prevWrongSubsets)$ \Comment{the count of every element in subsets of $X_{n}$ with sum $S$ called from function $GetED$ described in Algorithm \ref{Ch5AlgGetED}}
  	\State $p = SD[n][S]$ : number of subsets of $X_{n}$ with sum $S$
  	\item[]
  	
  \end{algorithmic}
  \begin{algorithmic}[1]
  	\Function{GenerateSubsets}{}
	  	\State $all\_buckets = p$ empty buckets
	  	\State \Call{initializeBuckets}{$all\_buckets$, $Table$, $n$, $S$, $p$}
	  	\Comment{Initial Step}
	  	\State $fillBuckets = True$ \Comment{Flag to control implementation of the while loop}
	  	\While {($fillBuckets$ is set \& ($|all\_buckets| > 0$))}
	  		\State $filBuckets = False$
	  		\State $q =$ count of non-zero entries of $Table$
	  		\State $x = min(p,q)$
	  		\State $elements = x$ largest integers of $X_{n}$ where $Table[ele] \neq 0$ $\forall \  ele \in elements$
	  		\State Sort $elements$ in descending order
	  		\For {$ele \in elements$}
	  			\State $b =$ \Call{findBucket}{$all\_buckets$, $ele$, $S$}
	  			\If {$b$ is a bucket} \Comment{When an elemnt can be inserted in a valid bucket.}
		  			\State Add $ele$ in bucket $b$
		  			\State $fillBuckets = True$ \Comment{If no element is alloted to any bucket in a full iteration.}
		  			\State $Table[ele]--$
		  			\If{Sum of the bucket $b == S$}
			  			\State $desiredSubsets += b$
			  			\State print bucket $b$ 
			  			\State Remove $b$ from $all\_buckets$ 
			  			\State $|all\_buckets|--$
		  			\EndIf
	  			\EndIf
	  		\EndFor
	  	\EndWhile
	  	\For {$bucket \in remaining\_buckets$}
	  		\State $wrongSubsets += bucket$
	  	\EndFor
	  	\State $Return \ wrongSubsets, \ Table$
	\EndFunction
  \end{algorithmic}
\end{algorithm}

\begin{algorithm}
  \caption{Basic BA: main Function($n$, $S$)}
  \label{Ch5AlgMainFunction}
  \begin{algorithmic}[1]
    \Function{mainFunction}{$n$, $S$}
    	\State $prevWrongSubsets = [ \ ]$
    	\State $prevTable = ED[n][S]$
    	\State $countSubsets = SD[n][S]$
    	\While{$countSubsets > 0$}
    		\State $prevWrongSubsets, prevTable = Generating Subsets(n, S, countSubsets$,
    		\item[]\hspace{\algorithmicindent} $\hfill prevTable, prevWrongSubsets)$
    		\State $countSubsets = SD[n][S] = |prevWrongSubsets|$ 
    		\item[]\hspace{\algorithmicindent} \Comment{Count of Subsets to be generated in next round is same as the size of wrong no. of subsets from previous round.}    	
    	\EndWhile
    	\State $Return  \ True$
    \EndFunction
  \end{algorithmic}
\end{algorithm}
For a given $n$ and $S$, time complexity of the algorithm depends on the maximum number of subsets and time to find a bucket for each element placement. Since, finding the bucket is an iterative algorithm, time taken for this sub-method is also proportional to the number of subsets, $SD[n][S]$. Since, the value of maximum number of subsets has exponential bound, $\mathcal{O}(2^n * n^{\frac{-3}{2}})$, as described in Appendix \ref{Chp5SumDistribution}, time complexity is $\mathcal{O}(max(SD[n][S] \cdot max(SD[n][S]) = \mathcal{O}(2^n * n^{\frac{-3}{2}} \cdot 2^n * n^{\frac{-3}{2}}) = \mathcal{O}(2^{2n} \cdot n^{-3})$. Therefore, given $n$ and $S$, the time complexity to generate all the subsets is $\mathcal{O}(2^{2n} \cdot n^{-3})$. Space complexity includes size of two storages $Table$ and $all\_buckets$, $\mathcal{O}(n) + \mathcal{O}(2^n) = \mathcal{O}(2^n)$.

\subsection{Subset Generation using Frequency Driven Bucket Algorithms}
\label{FrequencyDrivenBA}
After the basic bucket algorithm we present two more bucket algorithms. While the previous algorithm uses the direct information provided by element distribution, $ED[n][S][e]$ and sum distribution, $SD[n][S]$, in these two algorithms we use element distribution in decreasing or increasing order. In other words, instead of assigning elements to a corresponding bucket in descending order, we assign elements to buckets based on their frequencies. Frequency of an element in all the subsets of $X_{n}$ with sum $S$, by definition, is equal to the count of the elements, denoted by $ED[n][S][e]$. These algorithms are called \textit{Frequency-Driven (FD) Bucket} algorithms. These can be called minimum FD or maximum FD bucket algorithms.

Information used by these algorithms is same as the Basic Bucket Algorithm. While the basic bucket algorithm is iterative, the minimum or maximum frequency driven algorithms are recursive. Information required by this algorithm, properties of elements that should be followed and the measures by which we ensure uniqueness (i.e. using log and lookup tables) is same as the primitive algorithm defined in Section \ref{Ch5EnumerationTechnique1}.

Next, we generate all twenty subsets of $X_{10}$ with $Sum=15$. For both the algorithms, we select an element based on minimum or maximum frequency. In case of Minimum FD bucket algorithm, we select the maximum element with minimum frequency and recursively produce all the subsets of desired sum. For Maximum FD, we select maximum element with maximum frequency. In Table \ref{table:Ch5FDIterations=10S=15}, we log all the iterations for generating all twenty subsets of $X_{10}$ with $Sum=15$. Following points briefly describe the working of Minimum FD bucket algorithm:
\begin{enumerate}
\item By following the algorithm, we select element $10$. Since, $ED[10][15][10] = 3$, first iteration generates $3$ subsets: $\{\{10, 5\}, \{10, 4, 1\}, \{10, 3, 2\}\}$. This is shown in the first row of Table \ref{table:Ch5FDMinMaxSubsets}. All subsets are generated in eight iterations. 
\item In next three iterations, we choose elements-$9$, $8$ and $7$ respectively, to generate next thirteen subsets. This will results in production of sixteen subsets.
\item In every iteration we update the count of elements according to the resulting subsets.
\item In fifth iteration, we select element $2$ and recursively generate two subsets, $\{\{2, 6, 4, 3\}, \{2, 5, 4, 3, 1\} \}$. 
\end{enumerate}

For maximum frequency driven bucket algorithm we select the maximum element with maximum frequency in every iteration. 
\begin{enumerate}
\item All twenty desired subsets are produced in seven iterations.
\item Since, $ED[10][15][1] = ED[10][15][2] = 9$ and $max(1, 2)$, we select element $2$ and generate nine subsets.
\item In second iteration we select element $4$ and recursively generate next four subsets: $\{9, 6\}, \{9, 5, 1\}$, $\{9, 4, 2\}$ and $\{9, 3, 2, 1\}$.
\item Table \ref{table:Ch5FDIterations=10S=15} and Table \ref{table:Ch5FDMinMaxSubsets} presents the log entries and subsets corresponding to all iterations of maximum FD bucket algorithm for $X_{10}$ with $sum=15$.
\end{enumerate}   

\begin{table}
\footnotesize
\hspace{8 em}
\begin{tabular}[t]{|c|c|c|c|c|c|c|c|c|c|c|c|}
\hline
\parbox{2cm}{$Elements \  \rightarrow$ \\ $Iterations \downarrow $} & $1$ & $2$ & $3$ & $4$ & $5$ & $6$  & $7$ & $8$ & $9$ & $10$\\ \hline
$0^{th}\   iteration$ & 9 & 9 & 8 & 8 & 8 & 6 & 5 & 5 & 4 & \textbf{3}\\ \hline
$1^{st}\   iteration$ & 8 & 8 & 7 & 7 & 7 & 6 & 5 & 5 & \textbf{4} & 0\\\hline
$2^{nd}\   iteration$ & 6 & 6 & 6 & 6 & 6 & 5 & 5 & \textbf{5} & 0 & 0\\\hline
$3^{rd}\   iteration$ & 4 & 4 & 5 & 4 & 5 & 4 & \textbf{4} & 0 & 0 & 0\\\hline
$4^{th}\   iteration$ & 2 & \textbf{2} & 3 & 3 & 4 & 3 & 0 & 0 & 0 & 0\\ \hline
$5^{th}\   iteration$ & 2 & 1 & 2 & \textbf{2} & 3 & 3 & 0 & 0 & 0 & 0\\ \hline
$6^{th}\   iteration$ & 1 & 0 & 1 & 1 & 2 & \textbf{2} & 0 & 0 & 0 & 0\\ \hline
$7^{th}\   iteration$ & 1 & 0 & 1 & 0 & 1 & 1 & 0 & 0 & 0 & 0\\ \hline
\end{tabular}
\hspace{1 em}
\begin{tabular}[t]{|c|c|c|c|c|c|c|c|c|c|c|c|}
\hline
\parbox{2cm}{$Elements \  \rightarrow$ \\ $Iterations \downarrow $} & $1$ & $2$ & $3$ & $4$ & $5$ & $6$  & $7$ & $8$ & $9$ & $10$\\ \hline
$0^{th}\   iteration$ & 9 & \textbf{9} & 8 & 8 & 8 & 6 & 5 & 5 & 4 & 3\\ \hline
$1^{st}\   iteration$ & 5 & 0 & 4 & \textbf{4} & 5 & 4 & 3 & 2 & 2 & 2\\\hline
$2^{nd}\   iteration$ & \textbf{3} & 0 & 2 & 0 & 4 & 3 & 2 & 2 & 2 & 1\\\hline
$3^{rd}\   iteration$ & 0 & 0 & 1 & 0 & 2 & 1&  2 & 1 & 1 & \textbf{1}\\\hline
$4^{th}\   iteration$ & 0 & 0 & 1 & 0 & 1 & 1 & 2 & 1 & \textbf{1} & 0\\ \hline
$5^{th}\   iteration$ & 0 & 0 & 1 & 0 & 1 & 0 & 2 & \textbf{1} & 0 & 0\\ \hline
$6^{th}\   iteration$ & 0 & 0 & 1 & 0 & 1 & 0 & \textbf{1} & 0 & 0 & 0\\ \hline
$7^{th}\   iteration$ & 0 & 0 & 0 & 0 & 0 & 0 & 0 & 0 & 0 & 0\\ \hline
\end{tabular}
\hfill
\caption{Log table for iterations of Minimum and Maximum Frequency Driven Bucket Algorithm. We are generating all twenty subsets of $X_{10}$ with $Sum=15$. Every column denotes the frequency calculation for ten elements and every row denotes the frequency calculations in every iteration. In this table the frequency of every selected element in the previous iteration is marked as bold.}
\label{table:Ch5FDIterations=10S=15}
\end{table}

\begin{table}
\footnotesize
\begin{center}
\begin{tabular}{|c|c|c|c|}
\hline
$Iterations$ & $Selected \  Element$ & $Subsets$ & $|Subsets|$\\ \hline
$1^{st}\   iteration$ & $10$ & $\{\{10, 5\}, \{10, 4, 1\}, \{10, 3, 2\}\}$ & $3$\\ \hline
$2^{nd}\   iteration$ & $9$ & $\{\{9, 6\}, \{9, 5, 1\}, \{9, 4, 2\}, \{9, 3, 2, 1\}\}$ & $4$ \\ \hline
$3^{rd}\   iteration$ & $8$ & $\{\{8, 7\}, \{8, 6, 1\}, \{8, 5, 2\}, \{8, 3, 4\}, \{8, 4, 2, 1\}\}$ & $5$\\ \hline
$4^{th}\   iteration$ & $7$ & $\{\{7, 6, 2\}, \{7, 5, 3\}, \{7, 5, 2, 1\}, \{7, 4, 3, 1\}\}$ & $4$\\ \hline
$5^{th}\   iteration$ & $2$ & $\{\{2, 6, 4, 3\}, \{2, 5, 4, 3, 1\} \}$ & $2$ \\ \hline
$6^{th}\   iteration$ & $4$ & $\{\{4, 6, 5\}\}$ & $1$\\ \hline
$7^{th}\   iteration$ & $6$ & $\{\{6, 5, 3, 1\}\}$ & $1$\\ \hline
\multicolumn{3}{|c|}{$Total \  Number \ of  \ Subsets \rightarrow$} & $20$ \\ \hline 
\end{tabular}
\end{center}

\begin{center}
\begin{tabular}{|c|c|c|c|}
\hline
$Iterations$ & $Selected \  Element$ & $Subsets$ & $|Subsets|$\\ \hline
$1^{st}\   iteration$ & $2$ & \parbox{8cm}{\centering $\{\{2, 1, 3, 4, 5\}, \{2, 1, 3, 9\}, \{2, 1, 4, 8\}, \{2, 1, 5, 7\},$ \par $\{2, 3, 4, 6\}, \{2, 3, 10\}, \{2, 4, 9\}, \{2, 5, 8\}, \{2, 6, 7\}\}$} & $9$\\ \hline
$2^{nd}\   iteration$ & $4$ & $\{\{4, 1, 3, 7\}, \{4, 1, 10\}, \{4, 3, 8\}, \{4, 5, 6\}\}$ & $4$ \\ \hline
$3^{rd}\   iteration$ & $1$ & $\{\{1, 5, 9\}, \{1, 6, 8\}, \{1, 3, 5, 6\}\}$ & $3$\\ \hline
$4^{th}\   iteration$ & $10$ & $\{\{10, 5\}\}$ & $1$\\ \hline
$5^{th}\   iteration$ & $9$ & $\{\{9, 6\} \}$ & $1$ \\ \hline
$6^{th}\   iteration$ & $8$ & $\{\{8, 7\}\}$ & $1$\\ \hline
$7^{th}\   iteration$ & $7$ & $\{\{7, 5, 3\}\}$ & $1$\\ \hline
\multicolumn{3}{|c|}{$Total \  Number \ of  \ Subsets \rightarrow$} & $20$ \\ \hline 
\end{tabular}
\end{center}
\caption{Log table for iterations of Minimum and Maximum Frequency Driven Bucket Algorithm. We are generating all twenty subsets of $X_{10}$ with $Sum=15$. First column represent all the iterations, second column shows the chosen element as per the frequency. Third column stores the subsets and the fourth column denotes the count of these subsets.}
\label{table:Ch5FDMinMaxSubsets}
\end{table}

\subsection{Algorithm and Complexities}
\label{Ch5FDAlgoComplexity}
\noindent We state the pseudo codes for solving minimum and maximum FD bucket algorithms. Algorithm \ref{Ch5FDAlgGetED} updates the element distribution after every iteration and is called from Algorithm \ref{Ch5FDBucketAlgGenerateSubsets}. This update ensures that correct number of subsets are generated. In Line $5$, the element count is reduced according to the answer generated so far. The main function which was defined in Algorithm \ref{Ch5FDAlgMainFunction}, repeatedly calls the $GeneratingSubsetsbyFDBucketAlgo$ function and updates following information:
\begin{enumerate}
\item $countSubsets$ - No. of subsets left.
\item $fullTable$ - Element distribution of $X_{n}$ with $Sum=S$.
\item $elements$ - Remaining elements which form remaining subsets.
\item $elementsCovered$ - Elements which are not allowed or required to form remaining subsets.
\end{enumerate}

Apart from these helper methods, the main functionality is presented in Algorithm \ref{Ch5FDBucketAlgGenerateSubsets}. First we define the input for our algorithm. Line $2$ is the base case of our recursive algorithm. We terminate the recursion when the desired sum $S$. $S$ is less than zero or there are no $elements$ left to generate the subsets. In Line $7$ and Line $8$ we find $minKey$ and $minVal$ pair. In case of Minimum FD algorithm $(minKey, minVal)$ is the largest element with minimum frequency, $e \in [1, n]$ where $ED[n][S][e]$ is minimum. For maximum FD algorithm, we find $(maxKey, maxVal)$, the largest element with maximum frequency. The pseudo code for both algorithms are similar. Therefore, only minimum FD bucket algorithm is described. The main idea behind this algorithm is to find $minKey$, and generate subsets of $X_{n}$ with $Sum=[S-minKey]$. This means by adding $minKey$ to $elementsCovered$, in Line $10$, we do not include it in future partial subsets. In Line $11$, we recursively call $GeneratingSubsetsbyFDBucketAlgo$ function with modified values.
The remaining part of the code is divided in two conditions which are based on the return values from Line $11$. It can either be empty or non-empty. $minKey$ is appended to every returning subset of $desiredSubsets[S-minKey]$ and $elementsCovered$ are updated accordingly. In last few lines, we increase the count of $ED[n][S][e]$ for next iteration. This step ensures that the correct subsets are created in next iteration too.

For a given $n$ and $S$, time complexity of the maximum or minimum FD bucket algorithm depends on the maximum number of subsets and time taken to solve one recursion. Since, iterating through all elements is a recursive algorithm, time taken for this sub-method is also proportional to the number of subsets, $SD[n][S]$. Since, the value of maximum number of subsets has exponential bound, $\mathcal{O}(2^n * n^{\frac{-3}{2}})$, as described in Appendix \ref{Chp5SumDistribution}, time complexity is $\mathcal{O}(max(SD[n][S] \cdot max(SD[n][S]) = \mathcal{O}(2^n * n^{\frac{-3}{2}} \cdot 2^n * n^{\frac{-3}{2}}) = \mathcal{O}(2^{2n} \cdot n^{-3})$. Therefore, for given $n$ and $S$,the time complexity to generate all the subsets is $\mathcal{O}(2^{2n} \cdot n^{-3})$. Space complexity includes size of two storages $Table$ and $desiredSubsets$, $\mathcal{O}(n) + \mathcal{O}(2^n) = \mathcal{O}(2^n)$.

\begin{algorithm}
  \caption{Max FD: GetED($n , S, Table, desiredSubsets$)}
  \label{Ch5FDAlgGetED}
  \begin{algorithmic}[1]
	    \Function{GetED}{$n , S, Table, desiredSubsets$}
	    	\State $newTable = Table$
	    	\For {$subset \in desiredSubsets$}
	    		\For {$ele \in subset$}
	    			\State $newTable[ele] -= 1$ \Comment{Reducing count of elements according to $desiredSubsets$.}
	    		\EndFor
	    	\EndFor 
	    	\State $Return \ newTable$
	    \EndFunction
    \end{algorithmic}
\end{algorithm}

\begin{algorithm}
  \caption{Max FD: main Function($n$, $S$)}
  \label{Ch5FDAlgMainFunction}
  \begin{algorithmic}[1]
    \Function{mainFunction}{$n$, $S$}
    	\State $fullTable = ED[n][S]$ 
    	\State $countSubsets = SD[n][S]$
    	\State $elements = [1, 2 \ldots n]$ \Comment{Available elements}
    	\State $elementsCovered = [ \  ]$ \Comment{Elements covered so far}
    	\While{$countSubsets > 0$}
    		\State $desiredSubsets = Generating Subsets(n, S, countSubsets,$
    		\item[]\hspace{\algorithmicindent} $\hfill elements, elementsCovered, fullTable)$
    		\State $countSubsets = SD[n][S] - |desiredSubsets|$ 
    		\State Update $fullTable$, $elements$, $elementsCovered$ 
    		\item[]\hspace{\algorithmicindent} \Comment{Reduce frequency of elements according to $desiredSubsets$.}
    	  \EndWhile
    	\State $Return  \ True$
    \EndFunction
  \end{algorithmic}
\end{algorithm}

\begin{algorithm}
  \caption{Max FD: GeneratingSubsetsbyFDBucketAlgo($n, S, SD[n][S], elements, elementsCovered, ED[n][S]$)}
  \label{Ch5FDBucketAlgGenerateSubsets}
  \begin{algorithmic}[1]
  	\State Given: $n$, $S$, $SD[n][S]$ and $ED[n][S][i]$ where $i \in [1, n]$ 
  	\State $desiredSubsets = [ \  ]$ \Comment{$desiredSubsets$ are all the subsets of $X_{n}$ with sum $S$.}
  	\State $fullTable = GetED(n, S, ED[n][S], desiredSubsets)$ \Comment{the count of every element in subsets of $X_{n}$ with sum $S$ called from function $GetED$ described in Algorithm \ref{Ch5FDAlgGetED}}
  	\State $p = SD[n][S]$ : number of subsets of $X_{n}$ with sum $S$
  	\item[]
  	
  \end{algorithmic}
  \begin{algorithmic}[1]
  	\Function{GenerateSubsets}{}
  		\If {$S <= 0$ or $|elements| == 0$}
  			\State $Return \ desiredSubsets$
  		\EndIf
  		\State $countSubsets = SD[n]$
		\While {$countSubsets > 0$}
			\State $minVal = min(ED[n][S][e] > 0)$ 
			\State $minKey = max(e \ \forall e \in [1,n]  \ \& \  ED[n][S][e] == minVal)$
			\State $elements.remove(minKey)$
			\State $elementsCovered.add(minKey)$
			\State $desiredSubsets = Generating Subsets(n, S-minKey, countSubsets$
			    		\item[]\hspace{\algorithmicindent} $\hfill ,elements, elementsCovered, fullTable)$
			\If{$desiredSubsets[S-minKey]$ is empty}
				\State $countSubsets--$
				\State $ED[n][S][minKey]--$
				\State $desiredSubsets[S] = [[minKey]]$
				\State $print(desiredSubsets[S])$
				\State $elementsCovered.remove(minKey)$
			\Else
				\For {$A \in desiredSubsets[S-minKey]$}
					\State $ED[n][S][minKey]--$
					\If{$(minKey \notin A) \  \& \ (minKey+sum(A) == S)$}
						\If{$A.append(minKey) \ is \  unique$}
							\State $countSubsets--$
							\State $print(A)$
							\State $desiredSubsets[S].append(A)$
							\State $In \ elementsCovered \ add \ elements  \ of \ A$
						\EndIf
					\EndIf
				\EndFor
			\EndIf
	  	\EndWhile
		\For{$e \in ED[n][S][e] <= 0 $} 
			\State $elementsCovered.add(e)$
		\EndFor
		\For{$A \in desiredSubsets \ \& \ e \in A$}
			\State $ED[n][S][e]++$
		\EndFor
	  	\State $Return \ desiredSubsets$
	\EndFunction
  \end{algorithmic}
\end{algorithm}

\subsection{Subset Generation using Local Search}
\label{Chp5subsec:Local Search}
Our next enumeration technique for subset generation is called the Local Search. Before proceeding with this algorithm, we define two new types of subsets called Maximal and Minimal subsets. They act as the starting point for the local search algorithm.

\subsubsection{Maximal and Minimal Subsets}
\label{Chp5sec:Maximal and Minimal SubSets}
We present a new idea to categorize subsets of a given class. First, we divide the power set of $X_{n}$, $\mathcal P \left({X_{n}}\right)$, on the basis of their sum and then further partition these subsets according to their length. We have formulated and explained this selection process in Section \ref{Ch2:LengthSumDistribution1}.

For defining maximal subset we have the set of first $n$ natural numbers, $X_{n}$, sum($S$) which belongs to $[0, maxSum(n)]$ where $maxSum(n) = \frac{n(n+1)}{2}$ and length($l$) which belongs to $[0, n]$. Consider, $A$ denotes the subsets of $X_{n}$ with length $l$ and sum $S$. We denotes $A$ as $A = \{A_{1}, A_{2} \ldots A_{k}\}$ where $k = LD[n][S][l]$, the total count of subsets with length $l$ and sum $S$. $A_{i}$ represents $i^{th}$ subset of set $A$ and $A_{i, j}$ represents $j^{th}$ element of subset $A_{i}$ where $j \in [1, l]$. There exists a maximal subset of $X_{n}$ of length $l$ and sum $S$, $A_{maximal}$, is defined such that $\forall j \in [1, l] \  A_{maximal,j} > A_{p,j}$ where $p \in \{[1, k] - {maximal}\}$. There also exists a minimal subset, $A_{minimal}$, defined such that $\forall j \in [1, l] A_{minimal,j} > A_{p,j}$ where $p \in \{[1, k] - {minimal}\}$. 

The key point is that not all values of $A_{maximal,j}$ will be greater than $j^{th}$ element of other subsets in $A$ but there will surely be a subset for which first $q$ elements are greater than rest of the subsets, where $q \in [1, l]$. In order to generate the  subset $A_{maximal}$ for $X_{n}$ for a given sum $S$ and length $l$, we find the smallest possible element for every position, starting from the rightmost position. This pattern of element generation will ensure largest possible elements at the start of the subset, resulting in the maximal subset. Similarly, we find the largest possible element for every position of minimal subset starting from the rightmost position which ensures the smallest possible element at the start of the subset, resulting in the desired minimal subset. Table \ref{table:ExampleMaxMinSubset} displays the maximal and minimal subsets for every sum and length pair of $X_{5}$.
\bgroup
\def\arraystretch{1.5}
\begin{table}[htbp]
\begin{center}
\begin{tabular}{|c|c|c|c|c|}
\hline
$Sum$ & $Length$ & $Subsets$ & $Maximal Subset$ & $Minimal Subset$\\
\hline  \hline
$  0$ & $0$ & $\phi$ & $\phi$ & $\phi$\\ \hline
$  1$ & $1$ & $\{\{1\}\}$ & $\{1\}$ & $\{1\}$\\ \hline
$  2$ & $1$ & $\{\{2\}\}$ & $\{2\}$ & $\{2\}$\\ \hline
\multirow{2}{*}{$3$} & $1$ & $\{\{3\}\}$ & $\{3\}$ & $\{3\}$\\\cline{2-5}
& $2$ & $\{\{1, 2\}\}$ & $\{\{1, 2\}\}$ & $\{\{1, 2\}\}$\\ \hline
\multirow{2}{*}{$4$} & $1$ & $\{\{4\}\}$ & $\{4\}$ & $\{4\}$\\\cline{2-5}
 & $2$ & $\{\{1, 3\}\}$ & $\{\{1, 3\}\}$ & $\{\{1, 3\}\}$\\ \hline
\multirow{2}{*}{$5$} & $1$ & $\{\{5\}\}$ & $\{5\}$ & $\{5\}$\\\cline{2-5}
 & $2$ & $\{\{2, 3\}, \{1, 4\}\}$ & $\{2, 3\}$ & $\{1, 4\}\}$ \\ \hline
\multirow{2}{*}{$6$} & $2$ & $\{\{2, 4\}, \{1, 5\}\}$ & $\{2, 4\}$ & $\{1, 5\}$ \\\cline{2-5}
 & $3$ & $\{\{1, 2, 3\}\}$ & $\{1, 2, 3\}$ & $\{1, 2, 3\}$\\ \hline
\multirow{2}{*}{$7$} & $2$ & $\{\{3, 4\}, \{2, 5\}\}$ & $\{3, 4\}$ & $\{2, 5\}$ \\\cline{2-5}
 & $3$ & $\{\{1, 2, 4\}\}$ & $\{1, 2, 4\}$ & $\{1, 2, 4\}$\\ \hline
\multirow{2}{*}{$8$} & $2$ & $\{\{3, 5\}\}$ & $\{3, 5\}$ & $\{3, 5\}$\\\cline{2-5}
 & $3$ & $\{\{1, 3, 4\}, \{1, 2, 5\}\}$ & $\{1, 3, 4\}$ & $\{1, 2, 5\}$\\ \hline
\multirow{2}{*}{$9$} & $2$ & $\{\{4, 5\}\}$ & $\{4, 5\}$ & $\{4, 5\}$\\\cline{2-5}
 & $3$ & $\{\{2, 3, 4\}, \{1, 3, 5\}\}$ & $\{2, 3, 4\}$ & $\{1, 3, 5\}$\\ \hline
\multirow{2}{*}{$10$} & $3$ & $\{\{2, 3, 5\}, \{1, 4, 5\}\}$ & $\{2, 3, 5\}$ & $\{1, 4, 5\}$ \\\cline{2-5}
 & $4$ & $\{\{1, 2, 3, 4\}\}$ & $\{1, 2, 3, 4\}$ & $\{1, 2, 3, 4\}$\\ \hline
\multirow{2}{*}{$11$} & $3$ & $\{\{2, 4, 5\}\}$ & $\{2, 4, 5\}$ &  $\{2, 4, 5\}$\\\cline{2-5}
 & $4$ & $\{\{1, 2, 3, 5\}\}$ & $\{1, 2, 3, 5\}$ & $\{1, 2, 3, 5\}$\\ \hline
\multirow{2}{*}{$12$} & $3$ & $\{\{3, 4, 5\}\}$ & $\{3, 4, 5\}$ & $\{3, 4, 5\}$\\\cline{2-5}
 & $4$ & $\{\{1, 2, 4, 5\}\}$ & $\{1, 2, 4, 5\}$ & $\{1, 2, 4, 5\}$\\ \hline
$  13$ & $4$ & $\{\{1, 3, 4, 5\}\}$ & $\{1, 3, 4, 5\}$ & $\{1, 3, 4, 5\}$\\ \hline
$  14$ & $4$ & $\{\{2, 3, 4, 5\}\}$ & $\{2, 3, 4, 5\}$ & $\{2, 3, 4, 5\}$\\ \hline
$  15$ & $5$ & $\{\{1, 2, 3, 4, 5\}\}$ & $\{1, 2, 3, 4, 5\}$ & $\{1, 2, 3, 4, 5\}$\\ \hline
\end{tabular}
\end{center}
\caption{Maximal and minimal subsets for every sum and length pair of $X_{5}$.}
\label{table:ExampleMaxMinSubset}
\end{table}
\egroup

The core idea for the local search algorithm is to find all possible subsets of a particular length $l$ and sum $S$ where our starting subset can be a maximal or minimal subset. We find subsets by iterating over length between $l_{min}$ and $l_{max}$ where these are the minimum and maximum possible subsets of $X_{n}$ with sum $s$ respectively. This is a heuristic algorithm. Next, we present a few examples to explain local search using maximal and minimal subset respectively.

Maximal subset has the largest possible element at every position for a given sum $S$ and length $l$. Therefore, for local search starting with the maximal subset, we begin from left most element, decrement the first permissible element followed by increment of next permissible element. On contrary, minimal subset has smallest possible element at every position for a given sum $S$ and length $l$. Therefore, we begin from left most element, increment the first permissible element followed by decrement of next permissible element. Every increment or decrement consists of one unit.

\begin{figure}
\centering
\includegraphics[width=0.7\textwidth,height=0.8\textheight,keepaspectratio]{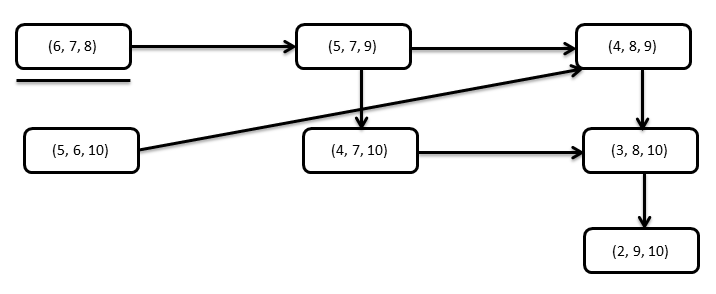}
\caption{Local search for $n = 10$, $sum = 21$ and $length = 3$ with maximal subset as the starting point.}
\label{Chp5fig:LocalSearchExample1}
\end{figure}

\begin{figure}
\centering
\includegraphics[width=0.7\textwidth,height=0.8\textheight,keepaspectratio]{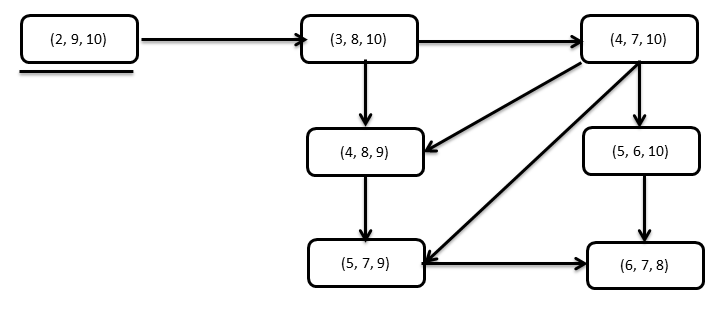}
\caption{Local search for $n = 10$, $sum = 21$ and $length = 3$ with minimal subset as the starting point.}
\label{Chp5fig:LocalSearchExample2}
\end{figure}

\begin{enumerate}
\item Figure \ref{Chp5fig:LocalSearchExample1} shows the local search example for $n = 10$, $sum = 21$ and $length = 3$ where the starting subset is the maximal subset of respective length. 
\begin{enumerate}
\item We start with subset $\{6, 7, 8\}$. By decrementing the first permissible element $6$ by $1$ and incrementing third permissible element $8$ by $1$, we generate the second subset $\{5, 7, 9\}$. We cannot increment the second element of subset $\{6, 7, 8\}$, as on incrementing $7$ by $1$, we get $8$ which creates duplication. In this case, $7$ is a non-permissible element.
\item Next, we generate subsets :$\{ \{4, 8, 9\}, \{5, 6, 10\}, \{4, 7, 10\} \}$ from subset $\{5, 7, 9\}$. 
\item By following the same procedure, we generate all desired subsets of $X_{10}$ with sum $21$ and length $3$ from a single maximal set $A_{maximal}$.
\end{enumerate}
\item Figure \ref{Chp5fig:LocalSearchExample2} presents the local search example for $n = 10$, $sum = 21$ and $length = 3$ where the starting subset is the minimal subset of respective length.
\begin{enumerate}
\item We start with subset $\{2, 9, 10\}$. By incrementing the first permissible element $2$ by $1$ and decrementing the second permissible element $9$ by $1$, we generate the second subset $\{3, 8, 10\}$. We can not decrement the third element of subset $\{2, 9, 10\}$, as on decreasing $10$ by $1$, we get $9$ which leads to duplication. In this case, $10$ is a non-permissible element.
\item Next, we generate subsets :$\{ \{4, 7, 10\}, \{4, 8, 9\}\}$ from subset $\{3, 8, 10\}$. 
\item By following the same procedure, we generate all desired subsets of $X_{10}$ with sum $21$ and length $3$ from a single minimal set, $A_{minimal}$.
\end{enumerate}
\item While generating a subset using Local Search Algorithm, we ensure that the sum of subset is equal to the desired target sum $S$, the subset do not contain duplicates and there is uniqueness among the subsets. Uniqueness among and within these subset is ensured by using lookup technique introduced in Section \ref{Ch5Lookup Table}. This establishes the correctness of the Local Search Algorithms using Maximal and Minimal Subsets.
\item Since we know the count of all subsets of $X_{n}$ with $Sum=S$ and $Length=l$, we generate all the subsets and this approach is concluded only when all desired subset results are achieved. This establish the completeness of the Local Search Algorithms using Maximal and Minimal Subsets.
\end{enumerate}

\noindent \textbf{Local Search using Maximal Subset:} Algorithm \ref{Chap5alg:LocalSearchMaximal} presents a procedure to generate all subsets of $X_{n}$ with particular sum $S$ and length $l$ where the seed subset is the maximal subset, $A_{maximal}$. We begin from the left most element, decrement the first permissible element followed by increment of next permissible element. Each increment or decrement consists of one unit. In Algorithm \ref{Chap5alg:LocalSearchMaximal}, we use a queue data structure to store all the resulting subsets, including maximal subset. We can iterate all the subsets in FCFS manner via these method. 
We check the uniqueness among the subsets by using the concept of lookup table as defined in Section \ref{Ch5Lookup Table}. A subset is pushed in the queue only if its unique. This algorithm is terminated when all the subsets are generated. \\

\noindent \textbf{Local Search using Minimal Subset:} Algorithm \ref{Chap5alg:LocalSearchMinimal} represents a procedure to generate all subsets of $X_{n}$ with particular sum $S$ and length $l$ where the seed subset is the minimal subset, $A_{minimal}$. We begin from left most element, increment the first permissible element followed by decrement of next permissible element. Every increment or decrement consists of one unit. Algorithm \ref{Chap5alg:LocalSearchMinimal} uses the same queue data structure and checks the uniqueness among the subsets by using the concept of lookup table as Algorithm \ref{Chap5alg:LocalSearchMaximal}. This algorithm is terminated when all subsets are generated. \\

\noindent \textbf{Complexities:} Time complexity of these algorithms is complexity of while loop $\times$ complexity of for loop, i.e., $maximum \  no. \  of \  subsets \  * \  length \  of \  each \  subset$. The complexity of the length of each subset variable is $\mathcal{O}(n)$ but the time complexity of $maximum \  no. \  of \  subsets$ variable is exponential. This makes the algorithm exhaustive. Time complexity is $\mathcal{O}(2^n \cdot n^{\frac{-3}{2}} \cdot n) = \mathcal{O}(2^n \cdot n^{\frac{-1}{2}}) = \mathcal{O}(\frac{2^n}{\sqrt{n}})$. The space complexity for these algorithms is equal to the size of storage queue i.e. $maximum \  no. \  of \  subsets \  * \  length \  of \  each \  subset$. The time complexity is similar. The complexity of the $\  Length \  of \  each \  subset$ variable is $\mathcal{O}(n)$ but the space complexity of $maximum \  no. \  of \  subsets$ variable is exponential, $\mathcal{O}(\frac{2^n}{\sqrt{n}})$.

\begin{algorithm}
  \caption{LS MaxS: Local Search for Maximal Subset}
  \label{Chap5alg:LocalSearchMaximal}
  \begin{algorithmic}[1]
  \Function{localSearch}{$n$, $s$, $l$}
    \State $maximalSet = $\Call{maximalSubset}{$n$, $s$, $l$}
    \State $queue.push(maximalSet)$
    \State $allSubsetsGenerated = LD[n][s][l]$
    \While {$allSubsetsGenerated > 0$}
    	\State $reqSet = queue.pop()$
    	\For {$i=1; i \leq len-1; i++$}
    		\If {$reqSet[i]-1 > reqSet[i-1]$}
    			\State $reqSet[i] -= 1$ \Comment{First decrementing the element by 1}
    			\State $decrement = True$
    		\EndIf
    		\For {$j=i+1; j \leq l; j++$}
    			\If {$reqSet[j]+1 < reqSet[j+1]$}
    				\State $reqSet[j]+=1$
	    			\State $increment = True$
    			\EndIf
    			\If {($reqSet$ is unique) and ($decrement$) and ($increment$)}
    				\State print $reqSet$
    				\State $queue.push(reqSet)$
    				\State $allSubsetsGenerated--$
    			\EndIf
    		\EndFor
    	\EndFor
    \EndWhile
  \EndFunction
  \end{algorithmic}
\end{algorithm}

\begin{algorithm}
  \caption{LS MinS: Local Search for Minimal Subset}
  \label{Chap5alg:LocalSearchMinimal}
  \begin{algorithmic}[1]
  \Function{localSearch}{$n$, $s$, $l$}
    \State $minimalSet = $\Call{minimalSubset}{$n$, $s$, $l$}
    \State $queue.push(minimalSet)$
    \State $allSubsetsGenerated = LD[n][l][s]$
    \While {$allSubsetsGenerated > 0$}
    	\State $reqSet = queue.pop()$
    	\For {$i=1; i \leq len-1; i++$}
    		\If {$reqSet[i]+1 < reqSet[i+1]$}
    			\State $reqSet[i] += 1$ \Comment{First incrementing the element by 1}
    			\State $increment = True$
    		\EndIf
    		\For {$j=i+1; j \leq l; j++$}
    			\If {$reqSet[j]-1 > reqSet[j-1]$}
    				\State $reqSet[j]-=1$
    				\State $deccrement = True$
    			\EndIf
    			\If {($reqSet$ is unique) and ($decrement$) and ($increment$)}
    				\State print $reqSet$
    				\State $queue.push(reqSet)$
    				\State $allSubsetsGenerated--$
    			\EndIf
    		\EndFor
    	\EndFor
    \EndWhile
  \EndFunction
  \end{algorithmic}
\end{algorithm}

\section{Experimental Results}
\label{Experimental Results}
This section presents the experiments that we have conducted to validate the efficiency and effectiveness of all the proposed algorithms.

\subsection{Summary of Alternate Enumeration Techniques}
\label{Ch:ConsolidatedSummaryConc}
Following table summarizes all the alternate enumeration techniques to solve SSP.

\begin{longtable}{| p{.16\textwidth} | p{.47\textwidth}| p{.18\textwidth}| p{.19\textwidth}|} 
\hline
\multicolumn{4}{|l|}{\textbf{Problem Statement:} Find all subsets of $\mathcal P \left({X_{n}}\right)$ which sum up to $S$, where $X_{n}$ is the set of first} \\
\multicolumn{4}{|l|}{ $n$ natural numbers, $X_{n} = \{1, 2 \ldots n\}$} \\ \hline
\textbf{Algorithm} & \textbf{Core Idea} & \textbf{Time Complexity} & \textbf{Space Complexity}\\ \hline 
Backtracking Algorithm (Naive) (section-\ref{Ch1Backtracking}) & It is an improved and systematic brute force approach for generating various subsets with $Sum=S$. We iterate through all $2^{n}$ solutions in an orderly fashion. & $\mathcal{O}(n \times 2^{n})$ & $\mathcal{O}(n)$\\
\hline
Subset Generator using Sum Distribution \newline (SDG) \newline (section-\ref{Ch5SumDisGen}) & This algorithm is a recursive generator based on the concept of Sum Distribution and uses subsets of $X_{(n-1)}$ to produce results for $X_{n}$. \newline Subsets of $X_{n}$ with $Sum=S$ are generated by subsets of $X_{n-1}$ with $Sum=(S-n)$. & $\mathcal{O}(2^n * n^{\frac{3}{2}})$ & $\mathcal{O}(2^n * n^{\frac{3}{2}})$ \\
\hline
Subset Generator using Length-Sum Distribution \newline (LDG) \newline (section-\ref{Ch5LengthSumDisGen}) & This algorithm is a recursive generator based on the concept of Length-Sum Distribution and uses subsets of $X_{(n-1)}$ to produce results for $X_{n}$. \newline Subsets of $X_{n}$ with $(Sum=S, Length=l)$ are generated by subsets of $X_{n-1}$ with $(Sum=S-n, Length=l-1)$. & $\mathcal{O}(2^n * n^{\frac{5}{2}})$ & $\mathcal{O}(2^n * n^{\frac{5}{2}})$ \\
\hline
Basic Bucket Algorithm \newline (Basic BA) \newline (section-\ref{Ch5EnumerationTechnique1}) & The basic idea behind this enumeration technique is to use the various distribution values. We consider $SD[n][S]$ number of empty buckets, storage data structures, and iterate through all elements in descending order. During each iteration an element is assigned to one of the buckets. This method is about adding the correct element to the corresponding subset. This is an iterative algorithm. & $\mathcal{O}(2^{2n} \cdot n^{-3})$ & $\mathcal{O}(2^n)$ \\
\hline
Maximum Frequency Driven Bucket Algorithm \newline (Max FD) \newline (section-\ref{FrequencyDrivenBA}) & Information used by this recursive algorithm is same as the basic bucket algorithm. Instead of choosing elements in descending order, we select maximum element with maximum frequency to generate all $SD[n][S]$ number of subsets of $X_{n}$ with $Sum=S$. & $\mathcal{O}(2^{2n} \cdot n^{-3})$ & $\mathcal{O}(2^n)$ \\
\hline
Minimum Frequency Driven Bucket Algorithm \newline (Min FD) \newline (section-\ref{FrequencyDrivenBA}) & This algorithm is contrary to maximum FD bucket algorithm. Information used by this is also similar to the basic bucket algorithm. We select maximum element with minimum frequency to generate all $SD[n][S]$ number of subsets of $X_{n}$ with $Sum=S$. & $\mathcal{O}(2^{2n} \cdot n^{-3})$ & $\mathcal{O}(2^n)$ \\
\hline
Local Search using Maximal Subset \newline (LS MaxS) \newline (section-\ref{Chp5subsec:Local Search}) & This heuristic algorithm finds all desired subsets by choosing the maximal subset as the seed. Maximal subset has the largest possible element at every position for a given sum($S$) and length($l$). Therefore, we begin from left most element, decrement the first permissible element followed by increment of next permissible element. Every increment or decrement consists of one unit. & $\mathcal{O}(\frac{2^n}{\sqrt{n}})$ & $\mathcal{O}(\frac{2^n}{\sqrt{n}})$ \\
\hline
Local Search using Minimal Subset \newline (LS MinS) \newline (section-\ref{Chp5subsec:Local Search}) & This heuristic algorithm also finds all desired subsets by choosing the minimal subset as the seed (starting point). Minimal subset has smallest possible element at every position for a given sum($S$) and length($l$). Therefore, we begin from left most element, increment the first permissible element followed by decremental of next permissible element. Every increment or decrement consists of one unit. & $\mathcal{O}(\frac{2^n}{\sqrt{n}})$ & $\mathcal{O}(\frac{2^n}{\sqrt{n}})$ \\
\hline
\caption{Summary of the core concepts and ideas of all the alternate enumeration techniques to solve subset sum problem. First column introduces every algorithm, second column presents the core idea behind the algorithm and the last two columns states their time and space complexities.}
\label{table:ConsolidatedSummary}
\end{longtable}

\subsection{Experimental Setup}
\label{Experimental Setup and Results}
We have carried out various sets of experiments on an i7-2600 machine with 64GB of RAM to compare and analyze the performance of our algorithms under various considerations. We define the experimental setup and measuring parameters before comparing the performances.  

\begin{figure}
\subfloat[Plot of number of subset of $X_{n}$ at $S=2n$ vs $n$]{\includegraphics[width = 3.4in]{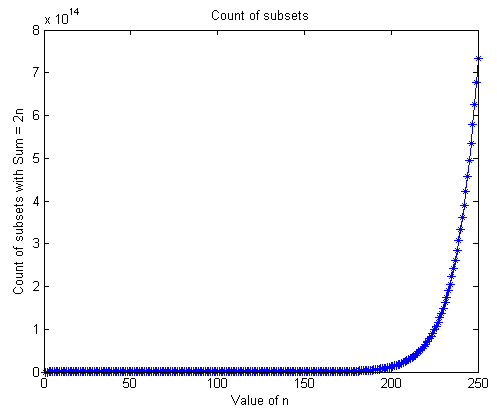}}
\subfloat[Plot of $log_{10}$(number of subset of $X_{n}$) at $2n$ vs $n$]{\includegraphics[width = 3.4in]{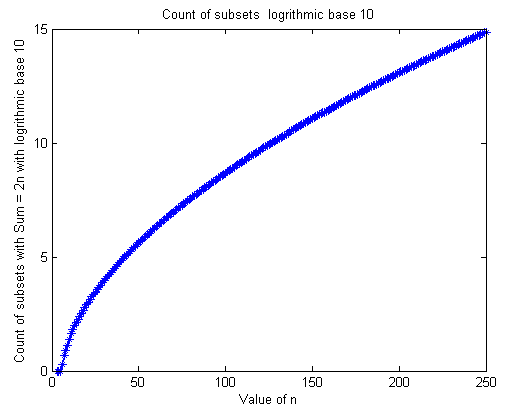}}\\
\subfloat[Plot of number of subset of $X_{n}$ at $S=(\frac{n(n+1)}{4}-n)$ vs $n$]{\includegraphics[width = 3.4in]{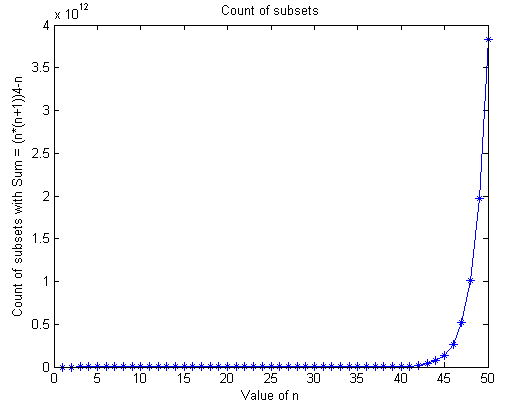}}
\subfloat[Plot of $log_{10}$(number of subset of $X_{n}$) at $S=(\frac{n(n+1)}{4}-n)$ vs $n$]{\includegraphics[width = 3.4in]{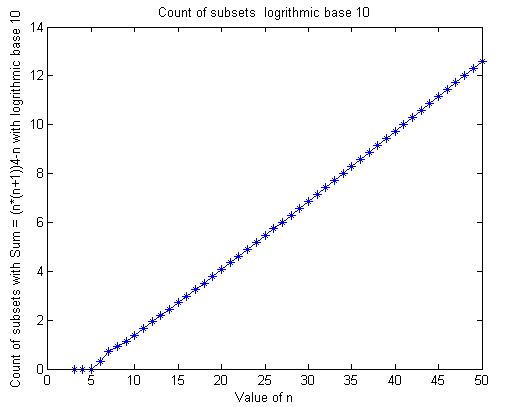}}
\caption{Plot of number of subsets of $X_{n}$ against sums in smaller and larger ranges. For smaller range we select $S=2n$ and plot graph for $n$ varying from $1$ to $250$. (a) Figure represents graph for number of subsets of $X_{n}$ with $S=2n$ where $n \in [1, 250]$. (b) Figure represents graph for number of subsets of $X_{n}$ with $S=2n$ with logarithmic base $10$ where $n \in [1, 250]$. For larger range we select $S=(\frac{n(n+1)}{4}-n)$ and plot graph for $n$ varying from $1$ to $50$. (c) Figure represents graph for number of subsets of $X_{n}$ with $S=(\frac{n(n+1)}{4}-n)$ where $n \in [1, 50]$. (d) Figure represents graph for number of subsets of $X_{n}$ with $S=(\frac{n(n+1)}{4}-n)$ with logarithmic base $10$ where $n \in [1, 50]$.}
\label{Results:SubsetCountComparision}
\end{figure}

Due to symmetric property of SSP, we choose random sum values in lower part of the sum range i.e. $S \leq midSum(n)$. In Figure \ref{Results:SubsetCountComparision}, we show different plots for number of subsets of $X_{n}$ for various sums. These figures help us estimate the problem space for generating results of alternate enumeration techniques. We select the value of $S$ as $2n$ to show the behavior of number of subsets of $X_{n}$ with sum $S$ when $S$ has the complexity $\mathcal{O}(n)$. Similarly, we choose the value of $S$ as $midSum(n)-n$ because the number of subsets of $X_{n}$ with this sum are in order of $\mathcal{O}(midSum(n))$. This upper bound of the Sum Distribution $SD[n][midSum(n)] = S(n) \approx \sqrt{\frac{6}{\pi}} \cdot 2^n \cdot n^\frac{-3}{2}$ is explained in Appendix \ref{Chp5SumDistribution}. Table \ref{table:ValuesofSubsets} presents the count of number of subsets $X_{n}$ with $S = 2n$ and $S = (\frac{n(n+1)}{4}-n)$. This table gives an estimate of the values plotted in Figure \ref{Results:SubsetCountComparision}. Figure \ref{Results:SubsetCountComparision}(a, c) plot the number of subsets of $X_{n}$ with $(n \in [1, 250], S=2n)$ and $(n \in [1, 50], S=(\frac{n(n+1)}{4}-n))$ respectively. Figure \ref{Results:SubsetCountComparision}(b, d) plot the log to the base $10$ of number of subsets of $X_{n}$ with $(n \in [1, 250], S=2n)$ and $(n \in [1, 50], S=(\frac{n(n+1)}{4}-n))$ respectively. Since the values of number of subsets for a particular $S$ increases exponentially with $n$, we have plotted Figure \ref{Results:SubsetCountComparision}(b, d) by using the logarithmic function. This helps in approximating the size of the problem space. 

\bgroup
\def\arraystretch{1.5}
\begin{table}[htbp]
\begin{center}
 \begin{tabular}{|c|c||c|c|}
\hline
$n$ & \parbox{1.3in}{\centering Count of Subsets \par of $X_{n}$ with \par $S=2n$} & $n$ & \parbox{1.3in}{\centering Count of Subsets \par of $X_{n}$ with \par $S=(\frac{n(n+1)}{4}-n)$} \\
\hline
\hline
6 & 2 & 6 & 2 \\ \hline
7 & 5 & 7 & 5\\ \hline
8 & 8 & 8 & 8\\ \hline
9 & 13 & 9 & 13\\ \hline
10 & 134 & 10 & 24\\ \hline
50 & 416868 & 15 & 521 \\ \hline
100 & 482240364 & 20 & 11812 \\ \hline
150 & 114613846376 & 30 & 7206286\\ \hline
200 & 11954655830925 & 40 & 5076120114\\ \hline
250 & 732839540340934 & 50 & 3831141038816\\
\hline
\end{tabular}
\end{center}
\caption{Count of number of subsets $X_{n}$ with $S = 2n$ and $S = (\frac{n(n+1)}{4}-n)$ respectively.}
\label{table:ValuesofSubsets}
\end{table}
\egroup

\subsection{Excess Subset Generation Analysis}
\label{Space Exploration}
Given $X_{n}$ and a sum $S$, we know how many subsets of $X_{n}$ have sum $S$. This value is $SD[n][S]$. For each algorithm, in order to generate these $SD[n][S]$ subsets we may explore few extra subsets of $X_{n}$ whose sum not equal to $S$. 

In naive backtracking method, at every step of subset generation we either include or exclude an element. This creates a recursive tree and a branch is terminated when the current sum exceeds the target. This way we explore more subsets than desired sum. Similarly, in rest of the alternate enumeration techniques in order to generate all subsets of $X_{n}$ with sum $S$, we explore more subsets than desired number of subsets. In this analysis we measure this extra exploration. In Table \ref{table:SpaceExploration} we present the ratios of subsets explored to total number of subsets of $X_{n}$ with sum $S$ i.e. $SD[n][S]$. The first three columns of this table states $(n, S)$ pair and the value of $SD[n][S]$ for all these pairs. The remaining eight columns denote the ratio of explored subsets to the number of subsets in the final solution for all eight alternate enumeration techniques. With every ratio we also represent the time taken by the algorithm to generate the solution set. For every value of $n$ and $S$, we bold the least ratio and least time taken by an algorithm.

Following observations can be made based on the data presented in Table \ref{table:SpaceExploration}:
\begin{enumerate}
\item For a given value of $n$ and $S$, desired ratio is a fraction of the number of subsets to be generated to the total number of subsets of $X_{n}$ with Sum $S$ i.e. $SD[n][S]$. For example, $n=12$ and $S=24$, the number of subsets of $X_{12}$ with $Sum=24$ are $67$. Therefore, the value of $SD[12][24] = 67$.
\item Every column corresponding to a given algorithm presents the ratio of number of subsets explored to generate the desired subsets to the total number of subsets of $X_{n}$ with sum $S$. For example for naive algorithm, given $n=12$ and $S=24$, the number of subsets explored for generating all subsets of $X_{12}$ with $Sum=24$ are $737$. Therefore, the desired ratio for these values is: $\frac{737}{67} = 11$.
\item The ratio for all algorithms should be greater than the desired ratio. If not, then it implies that complete result has not been generated. In this table for a given $n$ and $S$, the ratio for all algorithms is greater than the desired ratio. This observation and the correctness of these algorithms ensure the completeness of the results.
\item Naive algorithm explores most number of subsets in order to generate the desired subsets. Naive is the worst performing enumeration technique compares to all our proposed algorithms. This shows that all our alternate enumeration techniques perform better than the benchmark algorithm.
\item Ratios of LS MaxS and LS MinS are closer to the desired ratio for a given $n$ and $S$.
\begin{enumerate}
\item Since LS MaxS and LSMinS are heuristic algorithms, they explore lesser number of subsets as compared to naive algorithm.
\item For example, given $n=12$ and $S=24$, the number of subsets explored for LS MaxS and LS MinS are $93$ and $103$ respectively creating a ratio of $\frac{93}{67} = 1.3881$ and $\frac{103}{67} = 1.5373$.
\item The drawback for these algorithm is that they do not generate results for higher values of $n$ and $S$ within short amount of time. This is explained more in Section \ref{Experiments}.
\end{enumerate}
  
\item After Local Search algorithms, LDG and SDG are next in good performance ranking. Ratio for LDG is smaller and closer to desired ratio than SDG. 
\begin{enumerate}
\item Since LDG is a simple dynamic algorithm which generate the subsets based on their sum and length, it goes to one more level of categorization among subsets and minimizes the excess exploration of undesired subsets.
\item Given $n=12$ and $S=24$, the number of subsets explored by LDG are $150$ and ratio is $\frac{150}{67} = 2.2388$.
\item LDG has precedence over others as it can enumerate all subsets of $X_{n}$ for a considerable values of $n$ within short amount of time. The numbers are shown in Table \ref{table:SDGLDGNaiveComaprision} of Section \ref{Experiments}.
\item SDG explores more subsets than LDG but it performs better than naive. While naive implementation involves a recursive tree based on the inclusion and exclusion of an element at every step, SDG builds the subset by using the exact formula defined in Section \ref {Distribution Formulae}. 
\item Given $n=12$ and $S=24$, the number of subsets explored by SDG are $214$ and ratio is $\frac{214}{67} = 3.1940$.
\end{enumerate}
\item Performance of Max FD and Min FD is similar to SDG. For $n=12$ and $S=24$, MaxFD explores $166$ subsets and Min FD explores $241$ subsets. For other pairs of $n$ and $S$ these values are very close.
\item Basic Bucket algorithm (Basic BA) also performs better than naive but can not compute all subsets for a considerable value of $n$ and $S$ within short amount of time.
\end{enumerate}

\begin{longtable}[c]{| p{.03\textwidth} | p{.04\textwidth} ||p{0.1\textwidth}|| p{0.11\textwidth}| p{0.1\textwidth}| p{0.1\textwidth} || p{0.1\textwidth}| p{0.1\textwidth}| p{0.1\textwidth}||p{0.1\textwidth}|p{0.1\textwidth}|} 
\hline
$n$  &   $S$  &  $|Subsets|$ &  Naive &  SDG &  LDG &  Basic BA &  Max FD &  Min FD &  LS MaxS &  LS MinS \\ \hline
\hline
\multirow{2}{*}{$12$} &  \multirow{2}{*}{$24$} & \multirow{2}{*}{$67$} & 11 & 3.1940 & 2.2388 & 5.0896 & 2.4776 & 3.5970 & \textbf{1.3881} & 1.5373 \\ 
& & & (0.00247) & (0.009596) & \textbf{(0.00103)} & (1.618) & (0.195) & (1.822) & (0.019) & (0.016)  \\ \hline
\multirow{2}{*}{$12$} &  \multirow{2}{*}{$27$} & \multirow{2}{*}{$84$} & 20.5952 & 2.7857 & 2.0952 & 5.1548 & 2.6190 & 4.2262 & 1.3690 & \textbf{1.2976} \\ 
& & & (0.00178) & (0.012512) & \textbf{(0.00116)} & (2.394) & (0.405) & (2.958) & (0.02) & (0.024)\\ \hline
\multirow{2}{*}{$15$} &  \multirow{2}{*}{$30$} & \multirow{2}{*}{$186$} & 21.4194 & 6.2097 & 1.6882 & 4.7742 & 2.3871 & 4 & \textbf{1.1882} & 1.2903 \\ 
& & & (0.00795) & (0.014521) & \textbf{(0.00499)} & (21.208) & (1.381) & (7.641) & (0.11) & (0.133)\\ \hline
\multirow{2}{*}{$15$} &  \multirow{2}{*}{$45$} & \multirow{2}{*}{$521$} & 23.3282 & 2.8177 & 1.3013 &  \centering -  & 2.8503 & 1.3129 & 1.0211 & \textbf{1.0058} \\ 
& & & (0.01184) & (0.056801) & \textbf{(0.00472)} & \centering- & (14.031) & (353.746) & (0.798) & (0.955) \\ \hline
\multirow{2}{*}{$16$} &  \multirow{2}{*}{$32$} & \multirow{2}{*}{$253$} & 60.6087 & 8.2806 & 1.6719 & 5.4664 & 2.3478 & 3.7470 & \textbf{1.1621} & 1.2332 \\ 
& & & (0.01118) & (0.017711) & \textbf{(0.00615)} & (77.48) & (2.341) & (11.761) & (0.211) & (0.255)\\ \hline
\multirow{2}{*}{$17$} &  \multirow{2}{*}{$59$} & \multirow{2}{*}{$1764$} & 27.0947 & 2.9127 & 1.3622 &   \centering -  & 2.9892 &  \centering -  & \textbf{1.0176} & 1.1037 \\ 
& & & 0.03996) & (0.233748) & \textbf{(0.01556)} & \centering- & (153.31) & \centering- & (10.144) & (12.058)\\ \hline
\multirow{2}{*}{$20$} &  \multirow{2}{*}{$40$} & \multirow{2}{*}{$860$} & 73.6390 & 30.0447 & 1.8189 &  \centering -  & 2.2667 &  \centering -  & \textbf{1.0707} & 1.1191 \\ 
& & & (0.03277) & (0.055301) & \textbf{(0.04656)} & \centering- & (21.923) & \centering- & (2.81) & (3.438)\\ \hline
\multirow{2}{*}{$20$} &  \multirow{2}{*}{$85$} & \multirow{2}{*}{$11812$} &  28.4236 & 2.9261 & 1.6258 &  \centering -  &  \centering -  &  \centering -  & 1.2332 & \textbf{1.2281} \\ 
& & & (0.30453) & (2.08991) & \textbf{(0.1357)} & \centering- & (6981.574) & \centering- & (572.839) & (664.875)\\ \hline
\multirow{2}{*}{$22$} &  \multirow{2}{*}{$104$} & \multirow{2}{*}{$41552$} & 34.7394 & 2.9706 & \textbf{1.8080} & \centering - & \centering - & \centering-  & \centering- & -\\
& & & (1.21103) & (8.53939) & \textbf{(0.52493)} & \centering- & \centering- & \centering- & \centering- & - \\\hline
\caption{The ratios of subsets explored to total number of subsets of $X_{n}$ with sum $S$ i.e. $SD[n][S]$ is presented in this table.  The first three columns of this table states $(n, S)$ pair and the value of $SD[n][S]$ for all these pairs. The remaining eight columns denote subsets explored ratio for all eight alternate enumeration techniques. With every ratio we also represent the time taken by the algorithm to generate the solution set. The least ratio and least time taken for every $n$ and $S$ are presented in bold.}
\label{table:SpaceExploration}
\end{longtable}

\subsection{Comparative Analysis of Enumeration Algorithms}
\label{Experiments}
In this section, we present the time taken by various enumeration techniques under different conditions. Experiments defined in this section are categorized based on the range of input sum value corresponding to the set of natural numbers $X_{n}$. Given $X_{n}$, $Sum(A)$ belonging to the range $[0, maxSum(n)] = [0, \frac{n(n+1)}{2}]$ where $A \in \mathcal P \left({X_{n}}\right)$. Choosing different values of sum between $0$ to $maxSum(n)$ is the core idea behind these experiments. Table \ref{table:ExperimentSummary} summarizes the explanation of all these experiments.

\begin{longtable}[c]{| p{.23\textwidth} | p{.3\textwidth}|  p{.1\textwidth}|  p{.3\textwidth}|} 
\hline
Experiments / Comparative Analysis & Description and Examples & Algorithms & Tables or Figures\\ \hline \hline
CA-SSR $[1, 2n]$ & For this experiment we randomly choose sum $S_{1}$ from a smaller range and calculate the time taken to generate subsets of $X_{n}$ with $Sum = S_{1}$. For every values of $n$, this smaller range varies from $1$ to $2n$ i.e. $\forall n, \ S_{1} \in [1, 2n]$. & Basic BA, Max FD, Min FD, LS MaxS, LS MinS & Table \ref{table:LocalMaxMinSmallerRange}: Time taken (in seconds) by Basic BA, Max FD and Min FD in CA-SSR. \newline Table \ref {table:BasicAscDescEDSmallerRange}: Time taken (in seconds) by LS MaxS and LS MinS in CA-SSR.
\\ \hline
CA-LSR $[midSum(n)-n, midSum(n)]$ & For this experiment we randomly choose 
sum $S_{2}$ from a larger range and calculate the time taken by all the algorithms to generate subsets of $X_{n}$ with $Sum = S_{2}$. For every values of $n$, this larger range varies from $midSum(n)-n$ to $midSum(n)$ i.e. $\forall n \ S_{2} \in [midSum(n)-n, midSum(n)]$. & Basic BA, Max FD, Min FD, LS MaxS, LS MinS & Table \ref{table:BasicAscDescEDLargerRange}: Time taken (in seconds) by Basic BA, Max FD and Min FD in CA-LSR. \newline Table \ref{table:LocalMaxMinLargerRange}: Time taken (in seconds) by LS MaxS and LS MinS in CA-LSR. 
\\ \hline
CA-FSV & In this experiment instead of choosing random vales of $S$ for every algorithm against every $n$, we fix few pairs of $(n, S_{1})$ and $(n, S_{2})$ for all the algorithms where $S_{1} = 2*n$ and $S_{2} = midSum(n)-n$ & Basic BA, Max FD, Min FD, LS MaxS, LS MinS, LDG, SDG & Table \ref{table:FixedRangeValues}: presents the time taken by Max FD, Min FD, Basic BA, LS MaxS, LS MinS, LDG and SDG algorithms where $S_{1} = 2*n$ and $S_{2} = midSum(n)-n$
\\ \hline
CA-SLN & For this experiment instead of fixing the value of sum $S$, we vary $S$ from $0$ to $maxSum(n) = \frac{n(n+1)}{2}$. This experiment helps us in analyzing the performance of SDG and LDG algorithms against Naive (backtracking) algorithm. In this experiment we enumerate all $2^{n}$ subsets of $X_{n}$ & SDG, LDG, Naive & Table \ref{table:SDGLDGNaiveComaprision}: presents the comparison of SDG and LDG with naive backtracking algorithm. This table presents the time taken(in sec) while enumerating all $2^n$ subsets of $X_{n}$ for every value of sum $S$ in range $[0, \frac{n(n+1)}{2}]$. This is the time taken by these algorithms to enumerate each and every subset. \newline Figure \ref{fig:LDGSDGNaive}: Plot of SDG, LDG and Naive algorithm while enumerating all $2^n$ subsets of $X_{n}$ for every value of sum $S$ in range $[0, \frac{n(n+1)}{2}]$.
\\ \hline
\caption{Summary of the experimental setup for Comparative Analysis of Enumeration Algorithms. First column states the name, second columns describes the experiment, third column lists the algorithms for which the experiment is carried out and the fourth column presents the tables and figures stating the time taken by different algorithms under several conditions.}
\label{table:ExperimentSummary}
\end{longtable}

We have drawn these tables and shown these times for demonstrative purposes. We have observed the following by running all the eight algorithms:
\begin{enumerate}
\item From comparative analysis of algorithms in smaller range (CA-SSR) we can see that Basic BA, Max FD, Min FD, LS MaxS and LS MinS generate subsets till $n$ equal to $22$, $36$, $36$, $44$ and $44$ respectively and takes less than $35,000$ seconds.
\begin{itemize}
\item Since LS MaxS and LS MinS explores lesser number of extra subsets as shown in Section \ref{Space Exploration}, it takes lesser amount of time than bucket algorithms. 
\item Among these five algorithms, Basic BA explores maximum number of subsets, takes most time for execution and can generate results till smaller values of $n$.
\end{itemize}
\item Comparative analysis of algorithms in larger range (CA-LSR) follows similar pattern as CA-SSR. The value of sum selected in this range has higher value of $SD[n][S]$ which results in more execution time. LS MaxS and LS MinS perform the best in this experiment.
\item Comparative Analysis with Fixed Sum Values (CA-FSV) allows us to compare seven algorithms: Max FD, Min FD, Basic BA, LS MaxS, LS MinS, LDG and SDG for a fixed value of $n$ and $S$.
\begin{itemize}
\item From this comparative study, we can see that LDG and SDG outperforms all the other algorithms. They can be executed till $n=36$ and takes least amount of time.
\item Even though SDG and LDG explores more number of subsets, additional information required by these algorithms is much lesser than the additional information required by Local Search and Bucket Algorithms. 
\item SDG and LDG does not require the values of $SD[n][S]$ and $ED[n][S][e]$ at every step of execution. They do not need to maintain the current state of algorithm. This reduces the execution time.
\end{itemize}
\item From comparative analysis of SDG, LDG and Naive (CA-SLN) we compare LDG, SDG with naive to show that our alternate enumeration techniques performs better than the existing algorithms. Using naïve algorithm, we are not able to generate all the subset above $n$ equal to or greater than $24$. This limits the execution. But LDG and SDG can easily be computed till $n=34$ in less than $40$ minutes.
\end{enumerate}
These timings are implementation and machine dependent. The above results show that even though some algorithms explore fewer extra subsets but they take more time due to lack of efficient implementation, storage and memory constraint. 

\begin{table}[!htbp]
\footnotesize
\hspace{4em}
\begin{tabular}{|c|c|c||c|c|c|}
\hline
\multicolumn{6} {|c|} {Time taken(in sec) by LS MaxS in CA-SSR}\\ \hline
$n$ & $S$ & Time(in sec) & $n$ & $S$ & Time(in sec)\\ \hline\hline
3  & 1  & 0.00015  & 24  & 47  & 19.862\\ \hline
4  & 1  & 0.00012  & 25  & 36  & 2.0454\\ \hline
5  & 1  & 0.00015  & 26  & 10  & 0.0023\\ \hline
6  & 3  & 0.00024  & 27  & 17  & 0.0114\\ \hline
7  & 4  & 0.00023  & 28  & 5  & 0.002251\\ \hline
8  & 10  & 0.00098  & 29  & 8  & 0.002551\\ \hline
9  & 5  & 0.00039  & 30  & 8  & 0.00289\\ \hline
10  & 16  & 0.0037  & 31  & 53  & 168.747\\ \hline
11  & 20  & 0.0094  & 32  & 58  & 510.344\\ \hline
12  & 11  & 0.0018  & 33  & 31  & 1.00335\\ \hline
13  & 6  & 0.00070  & 34  & 56  & 411.957\\ \hline
14  & 1  & 0.00060  & 35  & 50  & 124.164\\ \hline
15  & 17  & 0.01081  & 36  & 47  & 67.4379\\ \hline
16  & 4  & 0.00086  & 37  & 62  & 1748.339\\ \hline
17  & 32  & 0.30375  & 38  & 74  & 18096.70\\ \hline
18  & 17  & 0.00682  & 39  & 56  & 588.9686\\ \hline
19  & 33  & 0.46984  & 40  & 58  & 951.2177\\ \hline
20  & 10  & 0.00144  & 41  & 32  & 1.890367\\ \hline
21  & 28  & 0.19325  & 42  & 55  & 561.6479\\ \hline
22  & 14  & 0.00403  & 43  & 46  & 76.02470\\ \hline
23  & 21  & 0.03176  & 44  & 44  & 48.61702\\ \hline
\end{tabular}
\hspace{3em}
\begin{tabular}{|c|c|c||c|c|c|}
\hline
\multicolumn{6} {|c|} {Time taken(in sec) by LS MinS in }\\ \hline
$n$ & $S$ & Time(in sec) & $n$ & $S$ & Time(in sec)\\ \hline\hline
3  & 1  & 0.00020  & 24  & 47  & 21.2260\\ \hline
4  & 1  & 0.00019  & 25  & 19  & 0.01893\\ \hline
5  & 1  & 0.00020  & 26  & 52  & 77.8854\\ \hline
6  & 2  & 0.00022  & 27  & 22  & 0.05621\\ \hline
7  & 7  & 0.00073  & 28  & 10  & 0.00278\\ \hline
8  & 6  & 0.00049  & 29  & 47  & 41.2712\\ \hline
9  & 9  & 0.00092  & 30  & 58  & 434.903\\ \hline
10  & 4  & 0.00047  & 31  & 10  & 0.00344\\ \hline
11  & 22  & 0.01392  & 32  & 12  & 0.00514\\ \hline
12  & 9  & 0.00122  & 33  & 10  & 0.00396\\ \hline
13  & 10  & 0.00145  & 34  & 27  & 0.356276\\ \hline
14  & 3  & 0.00069  & 35  & 43  & 26.97922\\ \hline
15  & 15  & 0.00634  & 36  & 44  & 36.25960\\ \hline
16  & 26  & 0.09706  & 37  & 32  & 1.74147\\ \hline
17  & 20  & 0.02005  & 38  & 34  & 3.16681\\ \hline
18  & 11  & 0.00148  & 39  & 59  & 1143.04\\ \hline
19  & 28  & 0.15588  & 40  & 13  & 0.00915\\ \hline
20  & 10  & 0.00149  & 41  & 55  & 556.808\\ \hline
21  & 31  & 0.43241  & 42  & 26  & 0.35234\\ \hline
22  & 28  & 0.22006  & 43  & 14  & 0.01238\\ \hline
23  & 20  & 0.02313  & 44  & 40  & 18.9293\\ \hline
\end{tabular}
\caption{Time taken (in seconds) by Local Search using Maximal Subset (LS MaxS) and Local Search using Minimal Subset (LS MinS) in CA-SSR where $S_{1}$ is randomly chosen  and $\forall n, \ S_{1} \in [1, 2n]$. \label{table:LocalMaxMinSmallerRange}}
\end{table}

\begin{table}[!htbp]
\footnotesize
\hspace{3em}
\begin{tabular}{|c|c|c|}
\hline
$n$ & $S$ & \parbox{2.5cm}{\centering Time taken(in sec) \par by Basic BA \par in CA-SSR}\\ \hline\hline
3 & 1 & 0.000551\\ \hline
4 & 1 & 0.0004.20\\ \hline
5 & 2 & 0.000138\\ \hline
6 & 3 & 0.000247\\ \hline
7 & 5 & 0.000405\\ \hline
8 & 2 & 0.000849\\ \hline
9 & 2 & 0.000885\\ \hline
10 & 16 & 0.05103\\ \hline
11 & 13 & 0.01842\\ \hline
12 & 13 & 0.02192\\ \hline
13 & 9 & 0.00265\\ \hline
14 & 16 & 0.08399\\ \hline
15 & 26 & 7.03249\\ \hline
16 & 31 & 60.6872\\ \hline
17 & 34 & 241.571\\ \hline
18 & 6 & 0.00106\\ \hline
19 & 19 & 0.75584\\ \hline
20 & 36 & 1261.39\\ \hline
21 & 15 & 0.09077\\ \hline
22 & 41 & 12918.6\\ \hline
\end{tabular}
\hfill
\hspace{1em}
\begin{tabular}{|c|c|c|}
\hline
$n$ & $S$ & \parbox{2.5cm}{\centering Time taken(in sec) \par by Max FD \par in CA-SSR}\\ \hline\hline
3 & 1 & 0.00076\\ \hline
4 & 1 & 0.000564\\ \hline
5 & 1 & 0.000569\\ \hline
6 & 3 & 0.001332\\ \hline
7 & 6 & 0.003052\\ \hline
8 & 10 & 0.000837\\ \hline
9 & 12 & 0.00139\\ \hline
10 & 7 & 0.00384\\ \hline
11 & 20 & 0.12669\\ \hline
12 & 24 & 0.19757\\ \hline
13 & 24 & 0.220844\\ \hline
14 & 5 & 0.0011010\\ \hline
15 & 2 & 0.0003778\\ \hline
16 & 9 & 0.0040118\\ \hline
17 & 11 & 0.007174\\ \hline
18 & 7 & 0.0024759\\ \hline
19 & 3 & 0.0008509\\ \hline
20 & 33 & 4.143428\\ \hline
21 & 11 & 0.007366\\ \hline
22 & 37 & 11.79708\\ \hline
23 & 29 & 1.868481\\ \hline
24 & 42 & 42.95121\\ \hline
25 & 22 & 0.274363\\ \hline
26 & 4 & 0.0009\\ \hline
27 & 5 & 0.001708\\ \hline
28 & 29 & 2.35588\\ \hline
29 & 27 & 1.51263\\ \hline
30 & 13 & 0.029837\\ \hline
31 & 15 & 0.068043\\ \hline
32 & 17 & 0.113950\\ \hline
33 & 57 & 1822.731\\ \hline
34 & 47 & 237.329\\ \hline
35 & 36 & 21.1548\\ \hline
36 & 71 & 32840.56\\ \hline
\end{tabular}
\hfill
\hspace{1em}
\begin{tabular}{|c|c|c|}
\hline
$n$ & $S$ & \parbox{2.5cm}{\centering Time taken(in sec) \par by Min FD \par in CA-SSR}\\ \hline\hline
3 & 1 & 0.00094\\ \hline
4 & 1 & 0.00065\\ \hline
5 & 2 & 0.00072\\ \hline
6 & 4 & 0.00156\\ \hline
7 & 1 & 0.00073\\ \hline
8 & 1 & 0.00073\\ \hline
9 & 3 & 0.00016\\ \hline
10 & 3 & 0.00172\\ \hline
11 & 7 & 0.00503\\ \hline
12 & 24 & 0.29195\\ \hline
13 & 21 & 0.14098\\ \hline
14 & 10 & 0.00586\\ \hline
15 & 23 & 0.25498\\ \hline
16 & 22 & 0.21270\\ \hline
17 & 34 & 4.35665\\ \hline
18 & 15 & 0.04607\\ \hline
19 & 20 & 0.15957\\ \hline
20 & 15 & 0.03706\\ \hline
21 & 9 & 0.00526\\ \hline
22 & 6 & 0.00236\\ \hline
23 & 34 & 8.08626\\ \hline
24 & 42 & 52.7242\\ \hline
25 & 9 & 0.00573\\ \hline
26 & 29 & 2.61190\\ \hline
27 & 33 & 8.01348\\ \hline
28 & 13 & 0.03223\\ \hline
29 & 51 & 523.152\\ \hline
30 & 24 & 0.73111\\ \hline
31 & 33 & 9.51999\\ \hline
32 & 41 & 71.6738\\ \hline
33 & 43 & 117.805\\ \hline
34 & 23 & 0.71141\\ \hline
35 & 14 & 0.08665\\ \hline
36 & 70 & 33113.13\\ \hline
\end{tabular}
\caption{Time taken (in seconds) by Basic Bucket Algorithm (Basic BA), Maximum Frequency Driven Bucket Algorithm (Max FD) and Minimum Frequency Driven Bucket Algorithm (Min FD) in CA-SSR where $S_{1}$ is randomly chosen  and $\forall n, \ S_{1} \in [1, 2n]$. \label{table:BasicAscDescEDSmallerRange}}
\end{table}

\begin{table}[!htbp]
\footnotesize
\begin{tabular}{|c|c|c|}
\hline
$n$ & $S$ & \parbox{2.5cm}{\centering Time taken(in sec) \par by Basic BA \par in CA-LSR}\\ \hline\hline
4 & 3 & 0.00079 \\ \hline
5 & 6 & 0.00102 \\ \hline
6 & 4 & 0.00044 \\ \hline
7 & 12 & 0.00882 \\ \hline
8 & 17 & 0.02263 \\ \hline
9 & 18 & 0.11157 \\ \hline
10 & 25 & 0.32170 \\ \hline
11 & 31 & 1.59752 \\ \hline
12 & 35 & 9.34307 \\ \hline
13 & 40 & 39.5506 \\ \hline
14 & 48 & 437.383 \\ \hline
15 & 50 & 1846.33\\ \hline
\end{tabular}
\hfill
\hspace{1em}
\begin{tabular}{|c|c|c|}
\hline
$n$ & $S$ & \parbox{2.5cm}{\centering Time taken(in sec) \par by Max FD \par in CA-LSR}\\ \hline\hline
3 & 3 & 0.00143\\ \hline
4 & 5 & 0.00153\\ \hline
5 & 7 & 0.00252\\ \hline
6 & 10 & 0.00522\\ \hline
7 & 14 & 0.01324\\ \hline
8 & 18 & 0.02944\\ \hline
9 & 22 & 0.07514\\ \hline
10 & 27 & 0.15790\\ \hline
11 & 33 & 0.41121\\ \hline
12 & 39 & 1.38157\\ \hline
13 & 45 & 4.16112\\ \hline
14 & 52 & 12.2192\\ \hline
15 & 60 & 39.9648\\ \hline
16 & 68 & 132.079\\ \hline
17 & 76 & 434.065\\ \hline
18 & 85 & 1426.70\\ \hline
19 & 95 & 4850.73\\ \hline
20 & 105 & 17189.86\\ \hline
\end{tabular}
\hfill
\hspace{1em}
\begin{tabular}{|c|c|c|}
\hline
$n$ & $S$ & \parbox{2.5cm}{\centering Time taken(in sec) \par by Min FD \par in Exp-$2$}\\ \hline\hline
3 & 2 & 0.00081\\ \hline
4 & 4 & 0.00129\\ \hline
5 & 3 & 0.00117\\ \hline
6 & 6 & 0.00329\\ \hline
7 & 7 & 0.00451\\ \hline
8 & 10 & 0.01452\\ \hline
9 & 13 & 0.03446\\ \hline
10 & 21 & 0.20658\\ \hline
11 & 31 & 1.87448\\ \hline
12 & 37 & 12.5400\\ \hline
13 & 33 & 7.19542\\ \hline
14 & 46 & 166.192\\ \hline
15 & 57 & 793.294\\ \hline
\end{tabular}
\caption{Time taken (in seconds) by Basic Bucket Algorithm (Basic BA), Maximum Frequency Driven Bucket Algorithm (Max FD) and Minimum Frequency Driven Bucket Algorithm (Min FD) in CA-LSR where $S_{2}$ is randomly chosen  and $\forall n, \ S_{2} \in [midSum(n)-n, midSum(n)]$. \label{table:BasicAscDescEDLargerRange}}
\end{table}

\begin{table}[!htbp]
\footnotesize
\hspace{4em}
\begin{tabular}{|c|c|c||c|c|c|}
\hline
\multicolumn{6} {|c|} {Time taken(in sec) by LS MaxS in CA-LSR}\\ \hline
$n$ & $S$ & Time(in sec) & $n$ & $S$ & Time(in sec)\\ \hline\hline
3 & 1 & 0.00003 & 13 & 39 & 0.12096\\ \hline
4 & 2 & 0.00025 & 14 & 49 & 0.34326\\ \hline
5 & 5 & 0.00057 & 15 & 47 & 0.77065\\ \hline
6 & 6 & 0.00049 & 16 & 54 & 2.65762\\ \hline
7 & 8 & 0.00067 & 17 & 72 & 10.4998\\ \hline
8 & 18 & 0.001429 & 18 & 67 & 30.6323\\ \hline
9 & 20 & 0.003262 & 19 & 87 & 149.328\\ \hline
10 & 18 & 0.00491 & 20 & 97 & 649.048\\ \hline
11 & 23 & 0.01419 & 21 & 94 & 1635.28\\ \hline
12 & 30 & 0.04604 & 22 & 114 & 8633.37\\ \hline
\end{tabular}
\hspace{2em}
\begin{tabular}{|c|c|c||c|c|c|}
\hline
\multicolumn{6} {|c|} {Time taken(in sec) by LS MinS in CA-LSR}\\ \hline
$n$ & $S$ & Time(in sec) & $n$ & $S$ & Time(in sec)\\ \hline\hline
3 & 1 & 0.00062 & 13 & 41 & 0.12125\\ \hline
4 & 2 & 0.00374 & 14 & 50 & 0.33774\\ \hline
5 & 6 & 0.00473 & 15 & 56 & 0.87691\\ \hline
6 & 7 & 0.01077 & 16 & 56 & 2.99347\\ \hline
7 & 11 & 0.00115 & 17 & 66 & 11.1003\\ \hline
8 & 13 & 0.0102 & 18 & 68 & 32.6975\\ \hline
9 & 14 & 0.00251 & 19 & 92 & 124.407\\ \hline
10 & 23 & 0.00898 & 20 & 100 & 639.238\\ \hline
11 & 29 & 0.02194 & 21 & 114 & 1881.25\\ \hline
12 & 31 & 0.05642 & 22 & 116 & 8897.909\\ \hline
\end{tabular}
\caption{Time taken (in seconds) by Local Search using Maximal Subset (LS MaxS) and Local Search using Minimal Subset (LS MinS) in CA-LSR where $S_{2}$ is randomly chosen  and $\forall n, \ S_{2} \in [midSum(n)-n, midSum(n)]$. \label{table:LocalMaxMinLargerRange}}
\end{table}

\FloatBarrier
\begin{longtable}[c]{| p{.03\textwidth} | p{.04\textwidth}| p{.085\textwidth}| p{.1\textwidth}| p{.08\textwidth} | p{.08\textwidth}| p{.08\textwidth}| p{.08\textwidth}|p{.08\textwidth}|p{.08\textwidth}|} 
\hline 
$n$ & $S$ & $|Subsets|$ & Max FD & Min FD & Basic BA & LS MaxS & LS MinS & LDG & SDG \\ \hline\hline
12 & 24 &  \centering67 & 0.195 & 1.822 & 1.618 & 0.019 & 0.016 & 0.00103 & 0.009596\\ \hline
12 & 27 &  \centering84 & 0.405 & 2.958 & 2.394 & 0.02 & 0.024 & 0.00116 & 0.012512\\ \hline
14 & 28 &  \centering134 & 0.808 & 3.95 & 14.76 & 0.066 & 0.065 & 0.00289 & 0.013285\\ \hline
14 & 38 &  \centering274 & 3.9 & 54.175 & 161.639 & 0.215 & 0.256 & 0.00315 & 0.03134\\ \hline
15 & 30 &  \centering186 & 1.381 & 7.641 & 21.208 & 0.11 & 0.133 & 0.00499 & 0.014521\\ \hline
15 & 45 &  \centering521 & 14.031 & 353.746 & 1388.5 & 0.798 & 0.955 & 0.00472 & 0.056801\\ \hline
16 & 32 &  \centering253 & 2.341 & 11.761 & 77.48 & 0.211 & 0.255 & 0.00615 & 0.017711\\ \hline
16 & 52 &  \centering965 & 45.83 & 1224.328 &  \centering -  & 2.882 & 3.394 & 0.0103 & 0.11678\\ \hline
17 & 34 &  \centering343 & 4.414 & 27.817 & 236.119 & 0.412 & 0.501 & 0.00821 & 0.024524\\ \hline
17 & 59 &  \centering1764 & 153.31 &  \centering -  &  \centering -   & 10.144 & 12.058 & 0.01556 & 0.233748\\ \hline
18 & 36 &  \centering461 & 7.913 & 52.816 & 649.679 & 0.791 & 0.96 & 0.02192 & 0.034023\\ \hline
18 & 67 &  \centering3301 & 541.046 &  \centering -  &  \centering -   & 39.129 & 45.385 & 0.03646 & 0.473109\\ \hline
20 & 40 &  \centering806 & 21.923 & 146.823 &  \centering -  & 2.81 & 3.438 & 0.04656 & 0.055301\\ \hline
20 & 85 &  \centering11812 & 6981.574 &  \centering -  &  \centering -   & 572.839 & 664.875 & 0.1357 & 2.08991\\ \hline
21 & 42 &  \centering1055 & 38.779 & 268.505 &  \centering -  & 5.177 & 6.298 & 0.0664 & 0.072871\\ \hline
21 & 94 &  \centering21985 & 25300.63 &  \centering -  &  \centering -   & 2084.648 & 2421.476 & 0.22368 & 4.134605\\ \hline
22 & 44 &  \centering1369 & 64.492 & 842.423 &  \centering -  & 9.411 & 11.516 & 0.09218 & 0.095904\\ \hline
22 & 104 &  \centering41552 &  \centering -  &  \centering -  &  \centering -  &  \centering -  &  \centering -  & 0.52493 & 8.53939\\ \hline
25 & 50 &  \centering2896 & 295.741 &  \centering -  &  \centering -   & 52.604 & 64.216 & 0.57455 & 0.211722\\ \hline
25 & 137 &  \centering283837 &  \centering -  &  \centering -  &  \centering -  &  \centering -  &  \centering -  & 2.35755 & 73.2227\\ \hline
27 & 54 &  \centering4649 & 831.93 &  \centering -  &  \centering -   & 155.258 & 190.273 & 1.06806 & 0.352428\\ \hline
27 & 162 &  \centering1038222 &  \centering -  &  \centering -  &  \centering -  &  \centering -  &  \centering -  & 10.77463 & 345.7571\\ \hline
30 & 60 &  \centering9141 &  \centering -  &  \centering -  &  \centering -   & 733.963 & 897.121 & 1.921185 & -\\ \hline
30 & 202 &  \centering7206286 &  \centering -  &  \centering -  &  \centering -  &  \centering -  &  \centering -  & 98.64595 & -\\
\hline\caption{The values of CA-FSV. We fix few pairs of $(n, S_{1})$ and $(n, S_{2})$ for Max FD, Min FD, Basic BA, LS MaxS, LS MinS, LDG and SDG algorithms where $S_{1} = 2*n$ and $S_{2} = midSum(n)-n$. This table shows time taken (in seconds) by all alternate techniques for calculating for these pairs.}
\label{table:FixedRangeValues}
\end{longtable}

\begin{longtable}{| p{.05\textwidth} | p{.15\textwidth}| p{.11\textwidth}| p{.11\textwidth}| p{.11\textwidth}|} 
\hline
$n$  & $|Subsets|$ & LDG & SDG & Naive \\\hline \hline
2 & 4 & 0.00030 & 0.00018 & 0.0008\\\hline
3 & 8 & 0.00024 & 0.00013 & 0.0012\\\hline
4 & 16 & 0.00032 & 0.00015 & 0.0014\\\hline
5 & 32 & 0.00037 & 0.00017 & 0.0026\\\hline
6 & 64 & 0.00053 & 0.00023 & 0.0054\\\hline
7 & 128 & 0.00061 & 0.00027 & 0.0116\\\hline
8 & 256 & 0.00099 & 0.00032 & 0.0218\\\hline
9 & 512 & 0.00122 & 0.00039 & 0.0423\\\hline
10 & 1024 & 0.00218 & 0.00052 & 0.0854\\\hline
11 & 2048 & 0.00257 & 0.00072 & 0.2137\\\hline
12 & 4096 & 0.00391 & 0.00103 & 0.4542\\\hline
13 & 8192 & 0.00532 & 0.00174 & 0.8522\\\hline
14 & 16384 & 0.00863 & 0.00272 & 1.5655\\\hline
15 & 32768 & 0.01223 & 0.00483 & 3.5153\\\hline
16 & 65536 & 0.02119 & 0.00927 & 6.1746\\\hline
17 & 131072 & 0.03060 & 0.01531 & 12.9676\\\hline
18 & 262144 & 0.05147 & 0.02669 & 24.3167\\\hline
19 & 524288 & 0.07002 & 0.05230 & 44.9257\\\hline
20 & 1048576 & 0.12088 & 0.10512 & 92.8140\\\hline
21 & 2097152 & 0.21260 & 0.20231 & 170.9037\\\hline
22 & 4194304 & 0.44724 & 0.39577 & 364.9816\\\hline
23 & 8388608 & 0.77863 & 0.81253 & 689.0156\\\hline
24 & 16777216 & 1.64562 & 1.60156 & - \\\hline
25 & 33554432 & 3.01883 & 3.13995 & - \\\hline
26 & 67108864 & 6.22996 & 6.21826 & - \\\hline
27 & 134217728 & 11.55410 & 12.60573 & - \\\hline
28 & 268435456 & 23.83728 & 25.29129 & - \\\hline
29 & 536870912 & 46.01338 & 49.41213 & - \\\hline
30 & 1073741824 & 97.38387 & 98.06444 & - \\\hline
31 & 2147483648 & 184.78691 & 202.59311 & - \\\hline
32 & 4294967296 & 375.63728 & 407.96308 & - \\\hline
33 & 8589934592 & 755.37561 & 844.82139 & - \\\hline
34 & 17179869184 & 2130.2298 & 2363.4442 & - \\\hline
\caption{Comparison of SDG and LDG with naive backtracking algorithm. This table presents the time taken(in sec) while enumerating all $2^n$ subsets of $X_{n}$ for every value of sum $S$ in range $[0, \frac{n(n+1)}{2}]$. This is the time taken by these algorithms to enumerate each and every subset in CA-SLN.}
\label{table:SDGLDGNaiveComaprision}
\end{longtable}

\begin{figure}[h!]
\begin{center}
\includegraphics{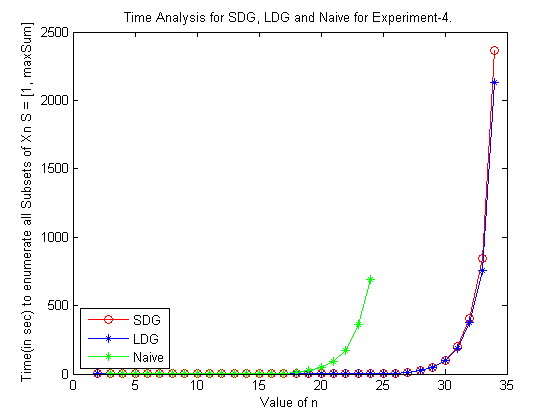}
\caption{Plot of SDG, LDG and Naive algorithm while enumerating all $2^n$ subsets of $X_{n}$ for every value of sum $S$ in range $[0, \frac{n(n+1)}{2}]$. This graph plots time taken by these algorithms to enumerate each and every subset in CA-SLN.}
\label{fig:LDGSDGNaive}
\end{center}
\end{figure}

\section{Conclusion}
\label{Conclusion}
Subset Sum Problem, also referred as SSP, is a well-known important problem in computing, cryptography and complexity theory. We extended the traditional SSP and suggested various alternate enumeration techniques. Instead of finding one subset with target sum, we find all possible solution of SSP. Therefore, for $X = \{5, 4, 9, 11\}$ and $S = 9$, the solution to our version of SSP is both $\{5, 4\}$ and $\{9\}$. We confined our problem domain by considering first n natural numbers as set $X_{n}$. In other words, we enumerate all $(2^{n} - 1)$ power set of a set.

We have analyzed the distribution of $\mathcal P \left({X_{n}}\right)$ over sum, length and count of individual elements. We introduced four types of distributions: Sum Distribution, Length Distribution, Length-Sum Distribution and Element Distribution. We extended the concept by explaining their formulae and algorithms, along with illustrations, which showed a definite pattern and relations among these subsets. These distributions are prepossessing procedures for various alternate enumeration techniques for solving SSP.

We developed Backtracking Algorithm (Naive) algorithm. It is an improved and systematic brute force approach for generating various subsets with $Sum=S$. Instead of searching exhaustively elements are selected systematically. We iterate through all $2^{n}$ solutions in this an orderly fashion. The inputs for this algorithm are the set of first $n$ natural numbers $X_{n}$ and $Sum = S$. Time and space complexities for this algorithm are $\mathcal{O}(n \times 2^{n})$ and $\mathcal{O}(n)$ respectively.  

We have proposed Subset Generator using Sum Distribution(SDG). This algorithm is a recursive generator based on the concept of Sum Distribution and uses subsets of $X_{(n-1)}$ to produce results for $X_{n}$. This algorithm uses the formula defined in Equation \ref{Ch2Eq1}. This algorithm is executed using dynamic programming. Subsets of $X_{n}$ with $Sum=S$ are generated by subsets of $X_{n-1}$ with $Sum=S$ and $Sum=(S-n)$. Time and space complexities for this algorithm are $\mathcal{O}(2^n * n^{\frac{3}{2}})$ and $\mathcal{O}(2^n * n^{\frac{3}{2}})$ respectively. 

We have proposed Subset Generator using Length-Sum Distribution (LDG). This algorithm is a recursive generator based on the concept of Length-Sum Distribution and uses subsets of $X_{(n-1)}$ to produce results for $X_{n}$. This algorithm uses the formula defined in Equation \ref{Ch2Eq12}. This algorithm is executed using dynamic programming. Subsets of $X_{n}$ with $(Sum=S, Length=l)$ are generated by subsets of $X_{n-1}$ with $(Sum=S, Length=l)$ and $(Sum=S-n, Length=l-1)$. Time and space complexities for this algorithm are $\mathcal{O}(2^n * n^{\frac{5}{2}})$ and $\mathcal{O}(2^n * n^{\frac{5}{2}})$ respectively. 

We have also proposed Basic Bucket Algorithm (Basic BA). The basic idea behind this enumeration technique is to use the various distribution values. We consider $SD[n][S]$ number of empty buckets, storage data structures, and iterate through all elements in descending order. It uses the value of Element Distribution for generating all the desired subsets. During each iteration an element is assigned to one of the buckets. This method is about adding the correct element to the corresponding subset. This is a greedy algorithm. This method uses the concept of lookup table explained in Section \ref{Ch5Lookup Table} and ensures uniqueness among and within the subsets. Time and space complexities for this algorithm are $\mathcal{O}(2^{2n} \cdot n^{-3})$ and $\mathcal{O}(2^n)$ respectively.

Next, we have extended the concept of Basic Bucket Algorithm (Basic BA) to propose two new bucket algorithms: Maximum Frequency Driven Bucket Algorithm (Max FD) and Minimum Frequency Driven Bucket Algorithm (Min FD). Information used by these recursive algorithms are same as the basic bucket algorithm. For Max FD, instead of choosing elements in descending order, we select maximum element with maximum frequency to generate all $SD[n][S]$ number of subsets of $X_{n}$ with $Sum=S$. For Min FD we select maximum element with minimum frequency to generate all $SD[n][S]$ number of subsets of $X_{n}$ with $Sum=S$. These methods use the concept of lookup table explained in Section \ref{Ch5Lookup Table} and ensure uniqueness among and within the subsets. Time and space complexities for this algorithm are $\mathcal{O}(2^{2n} \cdot n^{-3})$ and $\mathcal{O}(2^n)$ respectively. 

We have proposed two more algorithms Local Search using Maximal Subset (LS MaxS) and Local Search using Minimal Subset (LS MinS). Maximal and Minimal Subsets are a new idea for categorizing subsets of a given class. First, we divide the power set of $X_{n}$, $\mathcal P \left({X_{n}}\right)$, on the basis of their sum and then further partition these subsets according to their length. LS MaxS is a heuristic algorithm. It finds all the desired subsets by choosing the maximal subset as the seed. Maximal subset has largest possible element at every position for a given sum($S$) and length($l$). Therefore, we begin from left most element, decrement the first permissible element followed by increment of next permissible element. LS MinS is also a heuristic algorithm also finds all desired subsets by choosing the minimal subset as the seed. Minimal subset has the smallest possible element at every position for a given sum($S$) and length($l$). Therefore, we begin from left most element, increment the first permissible element followed by decremental of next permissible element. Every increment or decrement consists of one unit. Time and space complexities for this algorithm are $\mathcal{O}(\frac{2^n}{\sqrt{n}})$ and $\mathcal{O}(\frac{2^n}{\sqrt{n}})$ respectively.

\begin{conj}
There are algorithms that can enumerate all solutions of Subset Sum Problem for set $X_{n}$ and sum $S$ where $0 \leq S \leq \frac{n(n+1)}{2}$ with $\mathcal{O}(SD[n][S])$ complexity.
\end{conj}

\noindent An optimal algorithm should enumerate exactly $SD[n][S]$ subsets which are part of the solution.\\

\noindent This work can be extended in following ways:
\begin{enumerate}
\item By amortizing and combining different set of sums as one input set. Instead of running one sum at a time, we can group the sum values for running various alternate enumeration techniques. This will save the execution time by avoiding recalculations of subsets for smaller ranges.
\item Additionally, we can reduce the execution time of alternate enumeration techniques. These techniques are implementation and machine dependent. These timings are also data structure dependent. As part of future work, we would like to explore more data structures and more powerful machines to reduce the running times furthermore.
\item We have seen that the Local Search algorithm using Maximal or Minimal Subset comparatively explores less number of extra subsets and have better execution time than bucket algorithms. We can enhance this algorithm by using element distribution to limit the heuristic search, by finding different starting points and applying better distance formula for traversing through the solution space.
\end{enumerate}

\section*{Acknowledgement}
We thank Kannan Srinathan and Geeta Hooda for their discussion on this work.

\section*{Appendix}
\subsection*{Lookup Technique}
\label{Ch5Lookup Table}
Mapping of each subset with a unique integer is the basic concept used to define a lookup table for power sets of $X_{n}$, where $X_{n} = \{1, 2 \ldots n\}$. Lookup table ensures uniqueness among the subsets and within elements for a subset. This table helps us to maintain the uniqueness at runtime of any algorithm. This technique is implemented with the help of bit vectors. Bit vector is a compact data structure which hashes each subset $A = \{A_{1}, A_{2} \ldots A_{l}\}$ to the corresponding integer, denoted by $num$, $S_{num} = \sum_{i=1}^{l} 2^{A_{i}-1}$. We consider a hash of size $2^n$. This hash will maintain a one-to-one mapping between all the subsets of $X_{n}$ and is denoted by $\mathcal P \left({X_{n}}\right)$.

\subsection*{Upper Bound on Sum Distribution}
\label{Chp5SumDistribution}
In this section, we use definitions and formulas presented in \ref{Ch2sec:Formulation}. By using the maximum limit on the number of subsets with a particular sum, we find an upper bound of our problem.

$SD[n]$, defined in Section \ref{Ch2:SumDistribution} represents the count of all the subsets of $X_{n}$ divided over sum $S$ where $S \in [1, b]$ and $b = \frac{n(n+1)}{2}$ (Table \ref{Ch2Table:FormulaSummary}). The maximum value of $SD[n]$ is found at $midSum(n) = \floor{\frac{n(n+1)}{4}}$. Table \ref{table:n and sum_mid} represents the value of $SD[n][midSum(n)]$ for first $15$ natural numbers.

\bgroup
\def\arraystretch{1.5}
\begin{table}[htbp]
\begin{center}
\begin{tabular}{|c|c|c|c|c|c|c|c|c|c|c|c|c|c|c|c|c|c|}
\hline
$n$  & 1 & 2 & 3 & 4 & 5 & 6 & 7 & 8 & 9 & 10 & 11 & 12 & 13 & 14 & 15\\
\hline
$sd[n][midSum(n)]$ & 1 & 1 & 2 & 2 & 3 & 5 & 8 & 14 & 23 & 40 & 70 & 124 & 221 & 397 & 722 \\
\hline
\end{tabular}
\end{center}
\caption{Values of $SD[n][midSum(n)]$ for first 15 natural numbers}
\label{table:n and sum_mid}
\end{table}
\egroup

For each $n$, value of $SD[n][midSum(n)]$ presented in table \ref{table:n and sum_mid} is the coefficient of $x^\frac{n(n+1)}{4}$ in the expansion of $\{(1+x)(1+x^2)(1+x^3) \ldots (1+x^n)\}$. This coefficient is denoted as $S(n)$ and $S(n) \approx \sqrt{\frac{6}{\pi}} \cdot 2^n \cdot n^\frac{-3}{2}$ \cite{Authors23}. Therefore, value of maximum number of subsets with sum as $midSum(n)$ has exponential bound, $\mathcal{O}(2^n \cdot n^{\frac{-3}{2}})$. This result is vastly used throughout the thesis in order to find complexities of various enumeration techniques.

\FloatBarrier
\bibliographystyle{splncs03}
\bibliography{paper}

\end{document}